\pdfoutput=1
\documentclass[12pt,a4paper]{article}

\usepackage{subcaption}
\usepackage{ifthen} 
\usepackage{subcaption}
\newboolean{pdflatex}
\setboolean{pdflatex}{true} 

\newboolean{articletitles}
\setboolean{articletitles}{true} 

\newboolean{uprightparticles}
\setboolean{uprightparticles}{false} 

\def\paperauthors{LHCb collaboration} 
\def\paperasciititle{Search for the BsToMuMuGamma decay} 
\def\papertitle{Search for the \Bsmmg decay} 
\def\paperkeywords{{High Energy Physics}, {LHCb}} 
\def\papercopyright{\the\year\ CERN for the benefit of the LHCb collaboration} 
\def\paperlicence{CC BY 4.0 licence}
\def\paperlicenceurl{https://creativecommons.org/licenses/by/4.0/}

\usepackage[top=1in, bottom=1.25in, left=1in, right=1in]{geometry}

\usepackage{comment}

\usepackage{microtype}
\usepackage{lineno}  
\usepackage{xspace} 
\usepackage{caption} 

\usepackage{graphicx}  
\usepackage{color}
\usepackage{colortbl}
\graphicspath{{./figs/}} 

\usepackage{amsmath} 
\usepackage{amssymb}
\usepackage{amsfonts}
\usepackage{upgreek} 

\newcommand*\patchAmsMathEnvironmentForLineno[1]{%
\expandafter\let\csname old#1\expandafter\endcsname\csname #1\endcsname
\expandafter\let\csname oldend#1\expandafter\endcsname\csname
end#1\endcsname
 \renewenvironment{#1}%
   {\linenomath\csname old#1\endcsname}%
   {\csname oldend#1\endcsname\endlinenomath}%
}
\newcommand*\patchBothAmsMathEnvironmentsForLineno[1]{%
  \patchAmsMathEnvironmentForLineno{#1}%
  \patchAmsMathEnvironmentForLineno{#1*}%
}
\AtBeginDocument{%
\patchBothAmsMathEnvironmentsForLineno{equation}%
\patchBothAmsMathEnvironmentsForLineno{align}%
\patchBothAmsMathEnvironmentsForLineno{flalign}%
\patchBothAmsMathEnvironmentsForLineno{alignat}%
\patchBothAmsMathEnvironmentsForLineno{gather}%
\patchBothAmsMathEnvironmentsForLineno{multline}%
\patchBothAmsMathEnvironmentsForLineno{eqnarray}%
}

\usepackage{hyperxmp}

\usepackage[pdftex,
            pdfauthor={\paperauthors},
            pdftitle={\paperasciititle},
            pdfkeywords={\paperkeywords},
            pdfcopyright={Copyright (C) \papercopyright},
            pdflicenseurl={\paperlicenceurl},
            ]{hyperref}

\usepackage[bottom,flushmargin,hang,multiple]{footmisc}

\usepackage[all]{hypcap} 

\usepackage{xspace} 
\usepackage{upgreek}

\def\lhcb   {\mbox{LHCb}\xspace}

\def\babar  {\mbox{BaBar}\xspace}

\def\MagUp {\mbox{\em Mag\kern -0.05em Up}\xspace}

\ifthenelse{\boolean{uprightparticles}}%
{
 
 \def\Pgamma      {\ensuremath{\upgamma}\xspace}

 \def\Peta        {\ensuremath{\upeta}\xspace}

 \def\Pmu         {\ensuremath{\upmu}\xspace}                 
 \def\Pnu         {\ensuremath{\upnu}\xspace}                 
                  
 \def\Ppi         {\ensuremath{\uppi}\xspace}                 
                  
 \def\Prho        {\ensuremath{\uprho}\xspace}

 \def\Pphi        {\ensuremath{\upphi}\xspace}

 \def\Ppsi        {\ensuremath{\uppsi}\xspace}

 \def\PDelta      {\ensuremath{\Delta}\xspace}                 
 \def\PXi         {\ensuremath{\Xi}\xspace}                 
 \def\PLambda     {\ensuremath{\Lambda}\xspace}                 
 \def\PSigma      {\ensuremath{\Sigma}\xspace}                 
 \def\POmega      {\ensuremath{\Omega}\xspace}                 
 \def\PUpsilon    {\ensuremath{\Upsilon}\xspace}
 \let\oldPi\Pi
 \def\PPi         {\ensuremath{\oldPi}\xspace}

 \def\PB      {\ensuremath{\mathrm{B}}\xspace}                 
                  
 \def\PD      {\ensuremath{\mathrm{D}}\xspace}

 \def\PJ      {\ensuremath{\mathrm{J}}\xspace}                 
 \def\PK      {\ensuremath{\mathrm{K}}\xspace}

 \def\Pb      {\ensuremath{\mathrm{b}}\xspace}                 
 \def\Pc      {\ensuremath{\mathrm{c}}\xspace}                 
 \def\Pd      {\ensuremath{\mathrm{d}}\xspace}

 \def\Pi      {\ensuremath{\mathrm{i}}\xspace}

 \def\Ps      {\ensuremath{\mathrm{s}}\xspace}                 
 \def\Pt      {\ensuremath{\mathrm{t}}\xspace}

 \def\thebaroffset{0.0em}
}
{
 
 \def\Pgamma      {\ensuremath{\gamma}\xspace}

 \def\Peta        {\ensuremath{\eta}\xspace}

 \def\Pmu         {\ensuremath{\mu}\xspace}                 
 \def\Pnu         {\ensuremath{\nu}\xspace}                 
                  
 \def\Ppi         {\ensuremath{\pi}\xspace}                 
                  
 \def\Prho        {\ensuremath{\rho}\xspace}

 \def\Pphi        {\ensuremath{\phi}\xspace}

 \def\Ppsi        {\ensuremath{\psi}\xspace}                 
                  
 \mathchardef\PDelta="7101
 \mathchardef\PXi="7104
 \mathchardef\PLambda="7103
 \mathchardef\PSigma="7106
 \mathchardef\POmega="710A
 \mathchardef\PUpsilon="7107
 \mathchardef\PPi="7105
                  
 \def\PB      {\ensuremath{B}\xspace}                 
                  
 \def\PD      {\ensuremath{D}\xspace}

 \def\PJ      {\ensuremath{J}\xspace}                 
 \def\PK      {\ensuremath{K}\xspace}

 \def\Pb      {\ensuremath{b}\xspace}                 
 \def\Pc      {\ensuremath{c}\xspace}                 
 \def\Pd      {\ensuremath{d}\xspace}

 \def\Pi      {\ensuremath{i}\xspace}

 \def\Ps      {\ensuremath{s}\xspace}                 
 \def\Pt      {\ensuremath{t}\xspace}

 \def\thebaroffset{0.18em}
}
\newcommand{\offsetoverline}[2][\thebaroffset]{\kern #1\overline{\kern -#1 #2}}%

\makeatletter
\ifcase \@ptsize \relax
  \newcommand{\miniscule}{\@setfontsize\miniscule{4}{5}}
\or
  \newcommand{\miniscule}{\@setfontsize\miniscule{5}{6}}
\or
  \newcommand{\miniscule}{\@setfontsize\miniscule{5}{6}}
\fi
\makeatother

\DeclareRobustCommand{\optbar}[1]{\shortstack{{\miniscule (\rule[.5ex]{1.25em}{.18mm})}
  \\ [-.7ex] $#1$}}

\def\mup        {{\ensuremath{\Pmu^+}}\xspace}
\def\mun        {{\ensuremath{\Pmu^-}}\xspace} 

\def\neub       {{\ensuremath{\overline{\Pnu}}}\xspace}

\def\neumb      {{\ensuremath{\neub_\mu}}\xspace}

\def\g      {{\ensuremath{\Pgamma}}\xspace}

\def\dquark    {{\ensuremath{\Pd}}\xspace}

\def\squark    {{\ensuremath{\Ps}}\xspace}

\def\cquark    {{\ensuremath{\Pc}}\xspace}

\def\bquark    {{\ensuremath{\Pb}}\xspace}

\def\tquark    {{\ensuremath{\Pt}}\xspace}

\def\pion   {{\ensuremath{\Ppi}}\xspace}
\def\piz    {{\ensuremath{\pion^0}}\xspace}
\def\pip    {{\ensuremath{\pion^+}}\xspace}
\def\pim    {{\ensuremath{\pion^-}}\xspace}

\def\rhomeson {{\ensuremath{\Prho}}\xspace}
\def\rhoz     {{\ensuremath{\rhomeson^0}}\xspace}

\def\kaon    {{\ensuremath{\PK}}\xspace}

\def\KorKbar {\kern \thebaroffset\optbar{\kern -\thebaroffset \PK}{}\xspace}

\def\Kp      {{\ensuremath{\kaon^+}}\xspace}
\def\Km      {{\ensuremath{\kaon^-}}\xspace}

\def\Kstarz  {{\ensuremath{\kaon^{*0}}}\xspace}

\newcommand{\etaz}{\ensuremath{\Peta}\xspace}

\newcommand{\phiz}{\ensuremath{\Pphi}\xspace}

\def\D       {{\ensuremath{\PD}}\xspace}

\def\DorDbar {\kern \thebaroffset\optbar{\kern -\thebaroffset \PD}\xspace}
\def\Dz      {{\ensuremath{\D^0}}\xspace}

\def\Dp      {{\ensuremath{\D^+}}\xspace}
\def\Dm      {{\ensuremath{\D^-}}\xspace}

\def\DpDm    {\ensuremath{\Dp {\kern -0.16em \Dm}}\xspace}

\def\B       {{\ensuremath{\PB}}\xspace}

\def\BorBbar {\kern \thebaroffset\optbar{\kern -\thebaroffset \PB}\xspace}
\def\Bz      {{\ensuremath{\B^0}}\xspace}

\def\Bd      {{\ensuremath{\B^0}}\xspace}

\def\BdorBdbar {\kern \thebaroffset\optbar{\kern -\thebaroffset \Bd}\xspace}

\def\Bs      {{\ensuremath{\B^0_\squark}}\xspace}

\def\BsorBsbar {\kern \thebaroffset\optbar{\kern -\thebaroffset \Bs}\xspace}

\def\jpsi     {{\ensuremath{{\PJ\mskip -3mu/\mskip -2mu\Ppsi}}}\xspace}
\def\psitwos  {{\ensuremath{\Ppsi{(2S)}}}\xspace}

\def\Y#1S{\ensuremath{\PUpsilon{(#1S)}}\xspace}

\def\LorLbar     {\kern \thebaroffset\optbar{\kern -\thebaroffset \PLambda}\xspace}

\def\BF         {{\ensuremath{\mathcal{B}}}\xspace}
\def\BR         {\BF}

\newcommand{\decay}[2]{\ensuremath{#1\!\to #2}\xspace} 

\def\to                 {\ensuremath{\rightarrow}\xspace}

\def\CP                {{\ensuremath{C\!P}}\xspace}

\def\Vtd  {{\ensuremath{V_{\tquark\dquark}^{\phantom{\ast}}}}\xspace}

\def\Vts  {{\ensuremath{V_{\tquark\squark}^{\phantom{\ast}}}}\xspace}

\def\Bzero      {{\ensuremath{\B^0}}\xspace}

\def\Bmmg     {\decay{\Bzero}{\mup\mun\g}}

\def\Bmmpi    {\decay{\Bzero}{\mup\mun\piz}}

\def\BsPhiGam     {\decay{\Bs}{\phi \g}}
\def\Bsphig       {\BsPhiGam}

\def\BKstGam      {\decay{\Bzero}{\Kstarz \g}}

\def\Bsjpsie   {\decay{\Bs}{\jpsi\etaz}}

\def\AT#1     {\ensuremath{A_{\mathrm{T}}^{#1}}\xspace}           

\def\Bsmm     {\decay{\Bs}{\mup\mun}}

\def\Bsmmg   {\decay{\Bs}{\mup\mun\g}}
\def\C#1      {\ensuremath{\mathcal{C}_{#1}}\xspace}                       
\def\Cp#1     {\ensuremath{\mathcal{C}_{#1}^{'}}\xspace}                    
\def\Ceff#1   {\ensuremath{\mathcal{C}_{#1}^{\mathrm{(eff)}}}\xspace}        
\def\Cpeff#1  {\ensuremath{\mathcal{C}_{#1}^{'\mathrm{(eff)}}}\xspace}       
\def\Ope#1    {\ensuremath{\mathcal{O}_{#1}}\xspace}                       
\def\Opep#1   {\ensuremath{\mathcal{O}_{#1}^{'}}\xspace}                    

\newcommand{\nospaceunit}[1]{\ensuremath{\text{#1}}}       
\newcommand{\aunit}[1]{\ensuremath{\text{\,#1}}}       

\newcommand{\tev}{\aunit{Te\kern -0.1em V}\xspace}
\newcommand{\gev}{\aunit{Ge\kern -0.1em V}\xspace}
\newcommand{\mev}{\aunit{Me\kern -0.1em V}\xspace}
\newcommand{\kev}{\aunit{ke\kern -0.1em V}\xspace}
\newcommand{\ev}{\aunit{e\kern -0.1em V}\xspace}
 
\newcommand{\mevc}{\ensuremath{\aunit{Me\kern -0.1em V\!/}c}\xspace}
\newcommand{\gevc}{\ensuremath{\aunit{Ge\kern -0.1em V\!/}c}\xspace}
\newcommand{\mevcc}{\ensuremath{\aunit{Me\kern -0.1em V\!/}c^2}\xspace}
\newcommand{\gevcc}{\ensuremath{\aunit{Ge\kern -0.1em V\!/}c^2}\xspace}
\newcommand{\gevgevcccc}{\ensuremath{\gev^2\!/c^4}\xspace} 

\def\mum  {\ensuremath{\,\upmu\nospaceunit{m}}\xspace}

\def\fb   {\ensuremath{\aunit{fb}}\xspace}
\def\invfb   {\ensuremath{\fb^{-1}}\xspace}

\def\gsim{{~\raise.15em\hbox{$>$}\kern-.85em
          \lower.35em\hbox{$\sim$}~}\xspace}
\def\lsim{{~\raise.15em\hbox{$<$}\kern-.85em
          \lower.35em\hbox{$\sim$}~}\xspace}

\def\sPlot{\mbox{\em sPlot}\xspace}

\def\pt         {\ensuremath{p_{\mathrm{T}}}\xspace}

\def\ptot       {\ensuremath{p}\xspace}

\def\evtgen     {\mbox{\textsc{EvtGen}}\xspace}

\def\geant      {\mbox{\textsc{Geant4}}\xspace}

\def\photos     {\mbox{\textsc{Photos}}\xspace}

\def\pythia     {\mbox{\textsc{Pythia}}\xspace}

\def\tell1  {TELL1\xspace}
\def\ukl1   {UKL1\xspace}

\newcommand{\lhcborcid}[1]{\href{https://orcid.org/#1}{\hspace*{0.1em}\raisebox{-0.45ex}{\includegraphics[width=1em]{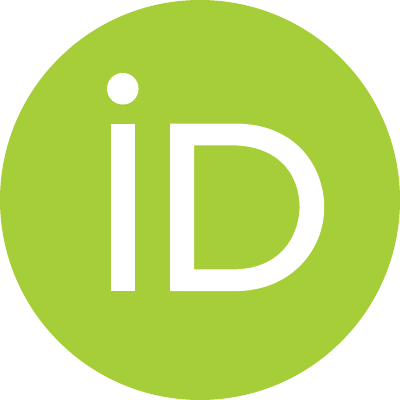}}}} 

\usepackage{cite} 
\usepackage{mciteplus}

\usepackage{longtable} 
\usepackage{multirow}
\usepackage{tabularx}

\usepackage[dvipsnames]{xcolor}
\hypersetup{
            colorlinks=true,
            linkcolor=violet,
            citecolor=teal,
            urlcolor=RoyalBlue
}

\definecolor{dartmouthgreen}{rgb}{0.05, 0.5, 0.06}

\begin{document}

\renewcommand{\thefootnote}{\fnsymbol{footnote}}
\setcounter{footnote}{1}


\begin{titlepage}
\pagenumbering{roman}

\vspace*{-1.5cm}
\centerline{\large EUROPEAN ORGANIZATION FOR NUCLEAR RESEARCH (CERN)}
\vspace*{1.5cm}
\noindent
\begin{tabular*}{\linewidth}{lc@{\extracolsep{\fill}}r@{\extracolsep{0pt}}}
\ifthenelse{\boolean{pdflatex}}
{\vspace*{-1.5cm}\mbox{\!\!\!\includegraphics[width=.14\textwidth]{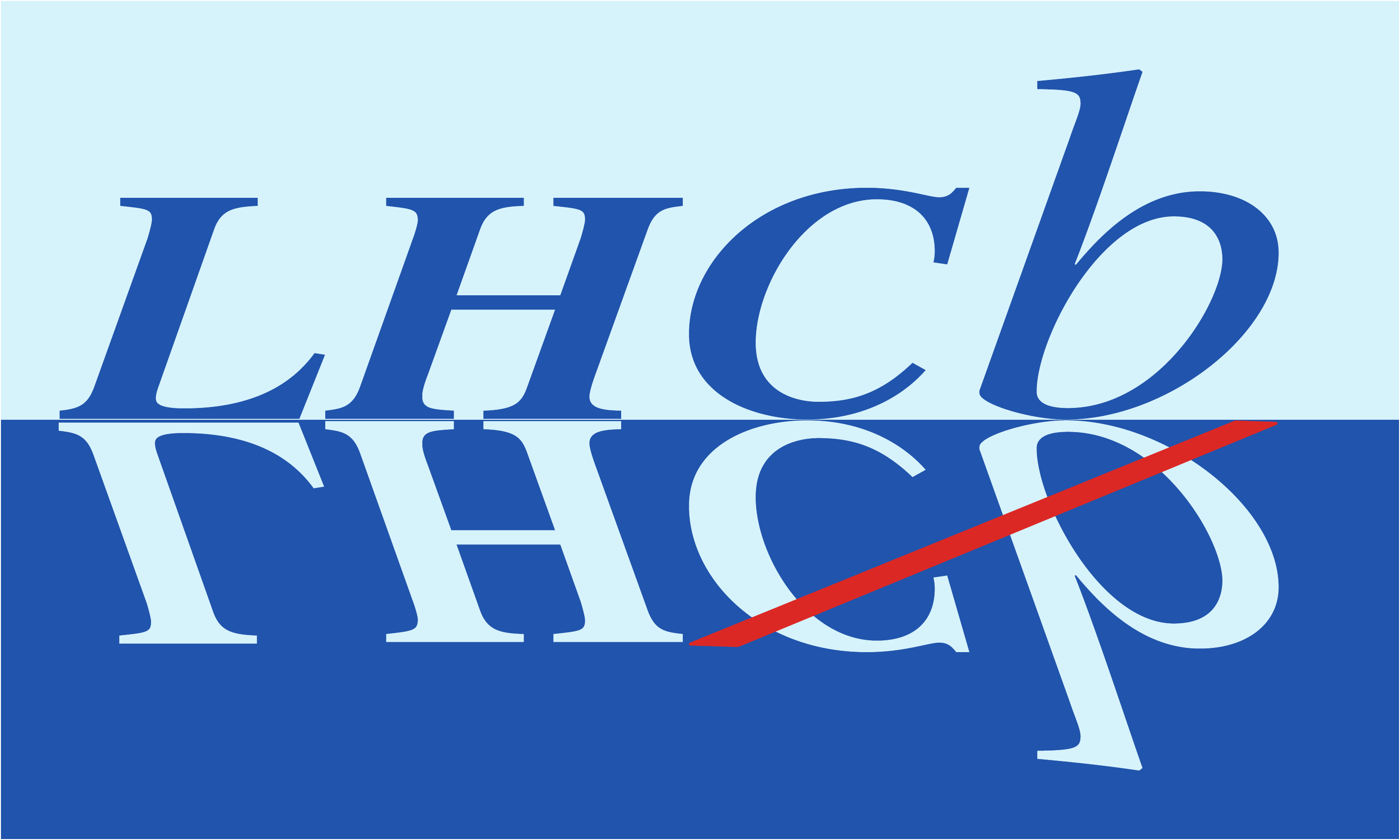}} & &}%
{\vspace*{-1.2cm}\mbox{\!\!\!\includegraphics[width=.12\textwidth]{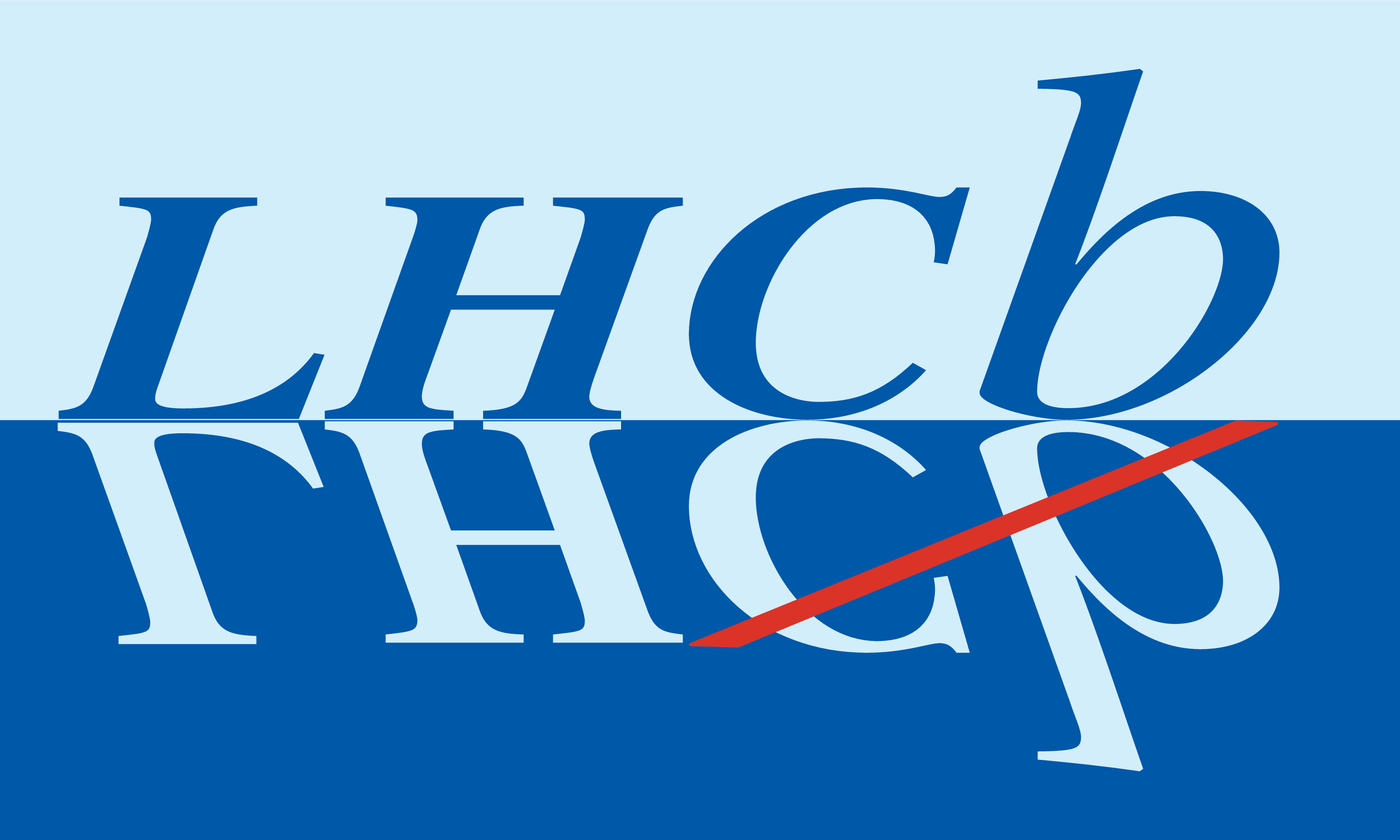}} & &}%
\\
 & & CERN-EP-2024-065 \\  
 & & LHCb-PAPER-2023-045 \\  
 & & April 4, 2024 \\ 
 & & \\
\end{tabular*}

\vspace*{4.0cm}

{\normalfont\bfseries\boldmath\huge
\begin{center}
  \papertitle 
\end{center}
}

\vspace*{2.0cm}

\begin{center}
\paperauthors\footnote{Authors are listed at the end of this paper.}
\end{center}

\vspace{\fill}

\begin{abstract}
  \noindent
  A search for the fully reconstructed \Bsmmg decay is performed at the LHCb experiment using proton-proton collisions at $\sqrt{s}=13\tev$ corresponding to an integrated luminosity of $5.4\invfb$. No significant signal is found and upper limits on the branching fraction in intervals of the dimuon mass are set
 \begin{align}
    \BR(\Bsmmg) < 4.2\times10^{-8},~&m(\mup\mun)\in[2m_\Pmu,~1.70]\gevcc ,\nonumber\\
    \BR(\Bsmmg) < 7.7\times10^{-8},~&m(\mup\mun)\in[1.70,~2.88]\gevcc ,\nonumber\\
    \BR(\Bsmmg) < 4.2\times10^{-8},~&m(\mup\mun)\in[3.92 ,~m_\Bs]\gevcc,\nonumber
\end{align}
at 95\% confidence level. Additionally, upper limits are set on the  branching fraction in the $[2m_\Pmu,~1.70]\gevcc$ dimuon mass region excluding the contribution from the intermediate $\phiz(1020)$ meson, and in the region combining all dimuon-mass intervals.

\end{abstract}

\vspace*{2.0cm}

\begin{center}
  Published in JHEP 07 (2024) 101
\end{center}

\vspace{\fill}

{\footnotesize 
\centerline{\copyright~\papercopyright. \href{\paperlicenceurl}{\paperlicence}.}}
\vspace*{2mm}

\end{titlepage}


\newpage
\setcounter{page}{2}
\mbox{~}
%
%
%
%


\renewcommand{\thefootnote}{\arabic{footnote}}
\setcounter{footnote}{0}

\cleardoublepage


\pagestyle{plain} 
\setcounter{page}{1}
\pagenumbering{arabic}


\section{Introduction}
\label{sec:Introduction}

Rare decays of \bquark hadrons involving flavour-changing neutral currents, such as those mediated by $b\to sll$ transitions, are forbidden at tree level in the Standard Model (SM) and are thus suppressed. As a consequence these decays are sensitive probes of potential contributions of beyond the SM (BSM) particles. Measurements of branching fractions and angular distributions
can probe new physics scenarios arising above the weak scale, encoded in the Wilson coefficients~\cite{Wilson_renormalization} of the weak effective theory, which at present are showing some tensions with the SM \cite{GlobalFit_2023}.

\begin{figure}[tb]
\centering
 \begin{subfigure}[t]{.45\textwidth}  
 \caption{} \includegraphics[height=3cm]{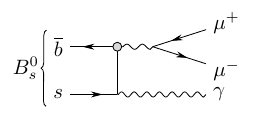}
 \end{subfigure}
 \begin{subfigure}[t]{.45\textwidth}
\caption{}\includegraphics[ height=3.2cm]{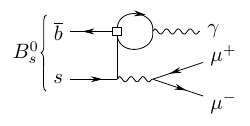}
 \end{subfigure}
\begin{subfigure}[t]{.45\textwidth}  
 \caption{} \includegraphics[height=3cm]{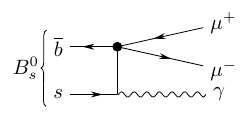}
    \end{subfigure}
\begin{subfigure}[t]{.45\textwidth} 
\caption{} \includegraphics[ height=2.9cm]{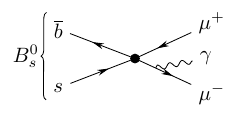} \end{subfigure} 
  \caption{ \small Diagrams contributing to \Bsmmg decay at lowest order.
    The grey circle denotes the electromagnetic-dipole operators $\mathcal{O}_7^{(\prime)}$. The empty square corresponds to any four-quark operator, where the quark-loop operators $\mathcal{O}_{1,2}$ dominate. The black circles denote the four-fermion operators $\mathcal{O}_{9,10}^{(\prime)}$. 
    }
  \label{fig:diagrams}
\end{figure}

\begin{figure}[tb]
    \centering\includegraphics[width=.75\textwidth]{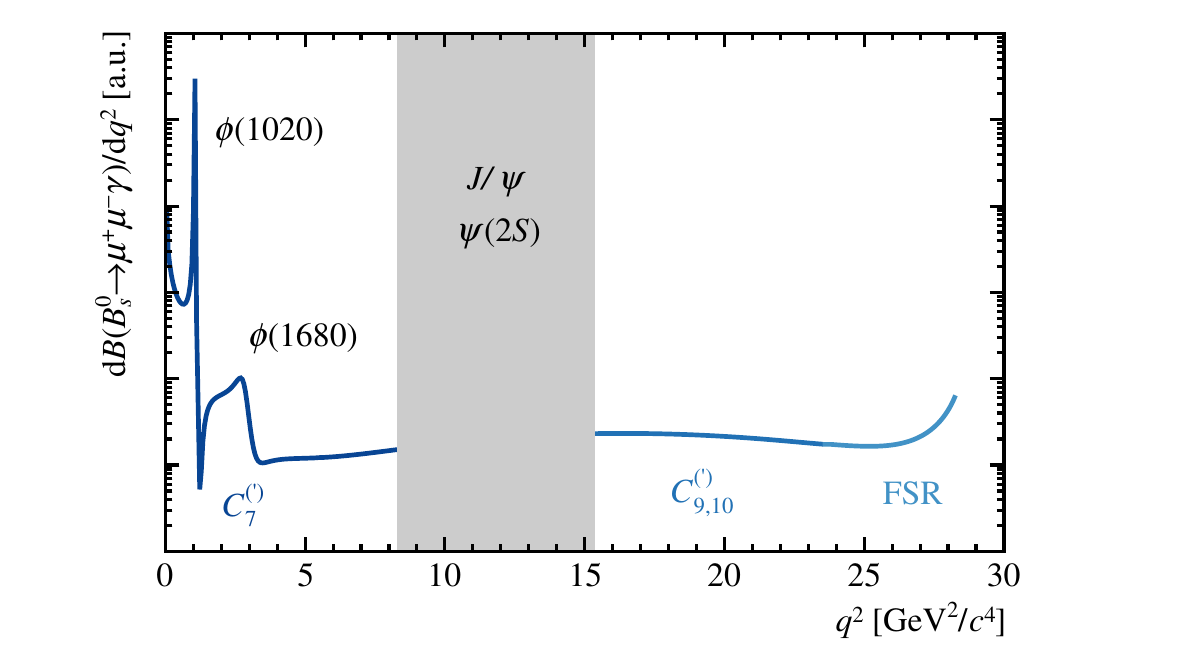}
    \caption{\small Illustration of the $q^2$ spectrum in \Bsmmg decays from Refs.~\cite{thesis_meril,MelikhovNikitin_2004} for a photon energy larger than 50\mevcc . The grey band corresponds to the excluded  region dominated by $\Bs\to\jpsi\gamma$ and $\Bs\to\psitwos\gamma$ decays. The Wilson coefficients and the FSR are highlighted in the region of the spectrum where they are dominant. The magnitudes are chosen for illustrative purposes.}
    \label{fig:q2spectrum}
\end{figure}

In this context, the \Bsmmg decay\footnote{The inclusion of charge-conjugate processes is implied throughout the text.} has attracted both theoretical and experimental interest as a powerful probe for investigating the aforementioned deviations from the SM. Compared to its nonradiative counterpart that is sensitive to the $\mathcal{O}_{10}^{(\prime)}$ operators, the \Bsmmg decay offers sensitivity to a wider set of operators~\cite{thesis_meril,BsToMuMuGamma_3,WilsonCoefs1,WilsonCoefs2}, as shown in Fig.~\ref{fig:diagrams}. Adding a photon to the dimuon final state lifts the chiral suppression factor present in the \Bsmm decay, enhancing the total branching fraction, compensating the addition of the QED vertex~\cite{theory4,theory2}.
However, this additional photon comes with the added challenge of the local form factors describing the $\Bs\to\gamma$ transitions.
At low dimuon mass squared, $q^2$, below the charmonium resonances,  the decay is mostly sensitive to the electromagnetic-dipole operators  $\mathcal{O}_7^{(\prime)}$. This kinematic region is especially interesting since different approaches to the calculation of the $\Bs\to\gamma$ local form factor result in different estimates of the branching fraction. In the low-$q^2$ region, these calculations have been done with single-pole~\cite{BsToMuMuGamma_Meril} and multipole~\cite{Nikitin_2018} parametrisations, soft-collinear effective theory (SCET)~\cite{Beneke_2020}, and using the light-cone sum rules (LCSR) approach~\cite{Janowski_2021}. Above the \psitwos resonance, this decay is sensitive to vector and axial-vector interactions dominated by the  $\mathcal{O}_{9,10}^{(\prime)}$ operators. In this high-$q^2$ region, the form factors are calculated with lattice QCD (LQCD), making use of heavy quark effective theory (HQET) extrapolation~\cite{frezzotti_2024} and assuming vector meson dominance (VMD)~\cite{Bsmumugamma_Camille,Bsmumugamma_Camille2}. 
Finally, at very high $q^2$ ($q^2 \gtrsim 25\gevgevcccc$), this decay is dominated by final state radiation (FSR). This variety of contributions shapes the \Bsmmg $q^2$ spectrum for a photon energy larger than 50\mevcc~\cite{thesis_meril,MelikhovNikitin_2004} as illustrated in Fig.~\ref{fig:q2spectrum}.
From Ref.~\cite{BsToMuMuGamma_Meril}, the SM predictions of the branching fractions are $\BF(\Bsmmg)_{\text{low-}q^2} = (8.3 \pm 1.3) \times 10^{-9}$ and $\BF(\Bsmmg)_{\text{high-}q^2} = (8.9 \pm 1.0) \times 10^{-10}$, where low- and high-$q^2$ regions are respectively defined by $q^2\in [0.04, 8.64] \gevgevcccc$ and $q^2 \in [15.84, 28.27] \gevgevcccc$.

The \babar collaboration published a first search for the \Bmmg decay, resulting in an upper limit of $\BF(\Bmmg)< 1.6 \times 10^{-7}$ at 90\% C.L~\cite{BaBar_B0}, while the \Bs counterpart was not explored.
At LHCb, the \Bsmmg decay was probed as a partially reconstructed background of the \Bsmm decay, and a first upper limit was set on the branching fraction: $\BF(\Bsmmg)< 2.0 \times 10^{-9}$ at 95\% C.L ~\cite{LHCb-PAPER-2021-008}.
This result is limited to the high-$q^2$ region, due to the restricted dimuon mass search window of [4.9\gevcc,~$m_{\Bs}$]. To be sensitive to the low-$q^2$ range, and therefore to a richer set of Wilson coefficients, the full final state reconstruction is needed.

This paper presents the first search for the \Bsmmg decay with full final state reconstruction, and the first search for this decay at low $q^2$.
Results are based on data collected with the LHCb detector in the years 2016\,--\,2018, corresponding to an integrated luminosity of 5.4\,fb$^{-1}$ of proton-proton ($pp$) collisions at a centre-of-mass energy $\sqrt{s}=13$\tev.


\section{Analysis strategy}
\label{sec:analysis}

The analysis is performed in the full kinematically accessible dimuon mass range, from the mass of the two muons, $2m_\Pmu$, up to the \Bs mass, $m_\Bs$. The analysis is performed in three dimuon mass bins as defined in Table~\ref{tab:BinFrac}. The value of 1.70\gevcc is chosen to be above the region where the $\phiz(1020)$\footnote{Referred to as $\phiz$ throughout the text.} and $\phiz(1680)$ resonances contribute. Using the SM prediction of the differential branching fraction derived in Ref.~\cite{BsToMuMuGamma_Meril}, the fraction of signal yield in each bin can be computed and is given in Table~\ref{tab:BinFrac}.

\begin{table}[tb]
\caption{\small Mass range definition, predicted branching fraction, and  the expected fraction of signal yield in the different $q^2$ bins as calculated in Ref.~\cite{BsToMuMuGamma_Meril}.}
    \centering \small
    \begin{tabular}{lccc}
        \hline\hline
        \rule{0pt}{2.5ex}$q^2$ bin~& I & II & III \\ 
        \hline 
        \rule{0pt}{2.5ex}$q^2$ [\gevgevcccc]& $[4m_\Pmu^2, 2.89]$ & $[2.89,8.29]$ & $[15.37,m_\Bs^2]$ \\
$m(\mup\mun)$ [\gevcc] & $[2m_\Pmu,1.70]$ & $[1.70,2.88]$ & $[3.92,m_\Bs]$ \\
        $10^{10}\times\BR(\Bsmmg)$& $82\pm15$ & $2.54\pm0.34$ & $9.1\pm1.1$\\
        Fraction of \Bsmmg  & 87\% & 2.7\% & 9.8\%\\
        \hline\hline
    \end{tabular}
    \label{tab:BinFrac}
\end{table}

Since bin~I is dominated by the \phiz resonance, a complementary study is done in bin~I excluding the \phiz mass region. This \phiz veto removes candidates with
\mbox{$m(\mup\mun)$ $\in$ [989.6,~1073.4]\mevcc}, as suggested in Ref.~\cite{BsToMuMuG_theo_2021}. The same selection optimisation and background modeling of bin~I are used for bin~I with the \phiz veto.

The signature of the \Bsmmg decay is two muons and one photon, with a combined invariant mass in the \Bs mass region 
and a decay vertex displaced with respect to the $pp$ interaction vertex.
This signature is exploited by the selection process, as described in Sec.~\ref{sec:selection}. It includes the description of the trigger, preselection, and the two multivariate classifiers based on a multilayer perceptron (MLP)~\cite{TMVA4} implemented in the TMVA toolkit~\cite{TMVA}, applied to reduce the backgrounds. The $\Bs\to\phiz\g$ decay, with $\phiz\to\Kp\Km$,  is used as control channel 
to assess the agreement of data and simulation as described in Sec.~\ref{subsec:control}.

The branching fraction of the signal is normalised to a well-known decay: $\Bs\to\jpsi\eta$, with $\jpsi\to\mup\mun$ and $\eta\to\g\g$ as presented in Sec.~\ref{sec:BRdetermination}. The signal yields are estimated using an extended unbinned maximum-likelihood fit to the final-state mass distribution, where the signal and background are modelled as described in Sec.~\ref{sec:model}. Finally the results are presented in Sec.~\ref{sec:results}.

To avoid experimenter's bias, the candidates
in the signal region, \mbox{$m(\mup\mun\g)\in[5.25,5.55]\gev/c^2$}, were not examined until the selection and analysis procedures were finalised. Most of the studies are done separately per data-taking year to account for possible variations in data-taking conditions, and merged at the last step of the analysis.


\section{Detector and simulation}
\label{sec:Detector}

The \lhcb detector~\cite{LHCb-DP-2008-001,LHCb-DP-2014-002} is a single-arm forward spectrometer covering the pseudorapidity range $2<\eta<5$, designed for the study of particles containing \bquark or \cquark quarks. The detector includes a high-precision tracking system consisting of a silicon-strip vertex detector surrounding the $pp$ interaction region~\cite{LHCb-DP-2014-001}, a large-area silicon-strip detector located upstream of a dipole magnet with a bending power of about $4{\mathrm{\,Tm}}$, and three stations of silicon-strip detectors and straw drift tubes~\cite{LHCb-DP-2017-001} placed downstream of the magnet.

The tracking system provides a measurement of the momentum, \ptot, of charged particles with a relative uncertainty that varies from 0.5\% at low momentum to 1.0\% at 200\gevc. The minimum distance of a track to a primary $pp$ collision vertex (PV), the impact parameter (IP), is measured with a resolution of $(15+29/\pt)\mum$ \cite{LHCb-DP-2014-001}, where \pt is the component of the momentum transverse to the beam, in \gevc.
Different types of charged hadrons are distinguished using information
from two ring-imaging Cherenkov detectors~\cite{LHCb-DP-2012-003}. 
Photons, electrons and hadrons are identified by a calorimeter system consisting of scintillating-pad (SPD) and preshower detectors (PRS), an electromagnetic (ECAL) and a hadronic (HCAL) calorimeter. Muons are identified by a
system composed of alternating layers of iron and multiwire
proportional chambers~\cite{LHCb-DP-2012-002}.

The online event selection is performed by a trigger~\cite{LHCb-DP-2012-004}, which consists of a hardware stage, based on information from the calorimeter and muon systems, followed by a software stage, which applies a full event reconstruction.

Simulation is used to optimise the selection strategy, to estimate the background and signal shapes, and to calculate the efficiencies. The $pp$ collisions are generated using \pythia~\cite{Sjostrand:2007gs,*Sjostrand:2006za} with a specific \lhcb configuration~\cite{LHCb-PROC-2010-056}. Decays of unstable particles are described by \evtgen~\cite{Lange:2001uf}, in which final-state radiation is generated using \photos~\cite{davidson2015photos}.  The interaction of the generated particles with the detector, and its response, are implemented using the \geant toolkit~\cite{Allison:2006ve, *Agostinelli:2002hh} as described in Ref.~\cite{LHCb-PROC-2011-006}. The theory model for the signal simulation is computed in Ref.~\cite{MelikhovNikitin_2004}.

In simulation, the same  reconstruction algorithms as in data are applied. Background candidates can contaminate the simulated samples if they are incorrectly reconstructed. A truth-matching algorithm is used to match reconstructed candidates with generated candidates. The algorithm matches the reconstructed objects with the generator level information for the final state particles and the particles in the decay chain.


\section{Selection}
\label{sec:selection}

Signal candidates are first selected by the hardware trigger, which requires candidates with either one muon with high transverse momentum, or one photon 
with high transverse momentum. Subsequently, the two-stage software trigger imposes requirements on the muon transverse momentum and track reconstruction quality.
To maximise the signal selection efficiency, candidates triggered by particles not associated with the signal candidates are also retained for further analysis, increasing the number of candidates by 10\%. A multivariate classifier based on topological criteria complements the software trigger selection~\cite{LHCb-PROC-2015-018}.
The same hardware and first-stage software trigger is used for the normalisation channel, the $\Bs\to\jpsi\eta$ decay. A second-stage software trigger that selects candidates with two reconstructed muons with invariant mass in the \jpsi mass region is used for the normalization channel.
The trigger for the control channel, $\Bs\to\phiz\g$, is chosen to select candidates with one photon and two charged tracks with reconstructed invariant mass in the \phiz mass region.

A preselection is applied to the \Bsmmg candidates.
The photon must have a transverse momentum larger than 1000\mevc. In order to separate photon clusters from non-electromagnetic charged clusters, a neural network classifier is employed~\cite{LHCb-DP-2020-001}. The input variables of this neural network are related to the energy deposits in each calorimeter and the shape of the cluster.
The muons must have a transverse momentum larger than 250\mevc and good track reconstruction quality. To avoid contributions from charmonia, candidates with a dimuon mass in the range $[2.88,3.92]$\gevcc are rejected. The \Bs candidate must have a transverse momentum larger than 500\mevc and a good quality decay vertex.

The preselection requirements on the normalisation channel are the same to that of the signal. Additional requirements on the \Bsjpsie candidates are that the two photons (muons) from the $\eta$ (\jpsi) decay must have an invariant mass within 105\mevcc (100\mevcc) of the known $\eta$ (\jpsi) mass~\cite{PDG2022}.

The candidates of the control channel are selected with the same requirements on the kinematic variables as the signal channel with some additional requirements to select \Bsphig candidates. The photon transverse momentum must be larger than 2.5\gevc and the invariant mass of the two kaons should be within 15\mevcc of the known \phiz mass~\cite{PDG2022}.

Further requirements on the particle identification (PID) information of the two muons are imposed in order to reject misidentified hadronic background. The muon identification (charged PID) uses multivariate techniques to combine information from different subsystems taking correlations into account~\cite{MuonID}.  The same muon identification requirement is applied to the signal and normalisation channels, while in the control channel, a different charged PID requirement is designed to select kaon tracks.


A further selection is applied to reject photons from $\piz\rightarrow\g\g$ decays where they are reconstructed as a single cluster due to the limited calorimeter granularity, referred to as merged. This photon identification is referred to as neutral PID. The neutral PID variable is the output of a multivariate classifier that is trained based on the shape of the electromagnetic cluster in the ECAL, the PRS and the SPD, for photons with
transverse momentum larger than 2\gevc . Photons with lower transverse momentum cannot be mistaken for merged neutral pions~\cite{PhotonReco}, and are therefore not subject to the neutral PID requirements. The same photon identification requirements are applied to the signal, normalisation and control channels.


The sample of selected candidates is dominated by random combinations of muons and photons from other \bquark-hadron decays in the same event, referred to as combinatorial background. To reduce this combinatorial contribution, candidates are rejected using a requirement on the response of an MLP classifier~\cite{MLP1,MLP2,MLP3}. 
This classifier is trained using  data outside the signal mass region defined in Sec.~\ref{sec:analysis} and simulated signal candidates weighted by the method explained in Sec.~\ref{subsec:control}. 
The variables used in the training are: the \Bs candidate vertex fit $\chi^2$, the candidate IP, and the IP significance;
 the cosine of the angle between the momentum vector of the \Bs candidate and the line joining the PV and its decay vertex, referred to as the direction angle; the minimum distance between the two muon tracks; the minimum IP of the muons with respect to any PV; the energy and transverse momentum of the photon; 
 the number of reconstructed \piz and $\eta$ mesons in the events that share a photon with the signal candidate; and the smallest mass difference between these \piz and $\eta$ candidates and their known masses.
 
 The performance of the MLP classifier is evaluated using the Punzi figure of merit (FoM) \cite{Punzi:2003bu} \begin{equation}\label{eq:FoM}
    \text{FoM} = \dfrac{\epsilon_\text{sig}}{\sigma/2 + \sqrt{N_{\text{bkg}}}} \,,
\end{equation}
where $\epsilon_\text{sig}$ is the signal efficiency, $\sigma = 5$ is the target statistical significance, and $N_{\text{bkg}}$ is the number of background candidates expected in the signal region.
The requirement on the MLP classifier is chosen to maximise the FoM of the signal, and the same requirement is applied to the control and normalisation channels.

To further reduce the remaining background, candidates are selected by a second MLP
classifier. It is trained separately on each $q^2$ region using simulated data as signal proxy and data outside the signal regions as background proxy. This second MLP classifier takes as input: the cosine of the angle between the momentum of the positive charged muon in the \Bs candidate rest frame and the vector perpendicular to the plane defined by the \Bs momentum and the beam axis; the \Bs candidate’s direction angle, its IP significance and its transverse momentum; the distance between the two muons in the $\etaz-\phi$ plane;\footnote{The symbols $\eta$ and $\phi$ here refer to the pseudorapidity and azimuthal angle (in radians).} the photon transverse momentum, energy and isolation; the muon smallest impact parameter significance, and the muon isolation.

The photon and muon isolation variables, quantify the probability that other tracks in the event originate from the same hadron decay as the signal candidates.

The photon isolation, $\text{I}(\g)$, is defined as
\begin{equation}
    \text{I}(\g) \equiv \frac{\pt(\g)}{\pt(\g) + \sum\limits_{\text{tracks}}\pt(\text{track})} ,
\end{equation}
where the sum in the denominator runs over all tracks within a cone of $\sqrt{\Delta \phi^2+\Delta \eta^2} = 1$ around the photon direction. Similar variables were used in previous measurements of radiative decays by \lhcb, as described in Refs.~\cite{LHCb-PAPER-2019-010,LHCb-PAPER-2021-017}.

Two isolation variables are designed to quantify the muon isolation, each considering a different type of track: tracks that have been reconstructed both before and after the magnet (long tracks) and tracks reconstructed only in the vertex detector (VELO tracks), as defined in Ref.~\cite{LHCb-PAPER-2021-008}.

The performance of the second MLP is also evaluated using the FoM from Eq.~\ref{eq:FoM}.
The optimal MLP requirement is chosen to maximise the FoM of the signal individually in each $q^2$ region. A loose MLP requirement is applied to the normalization channel in order to keep a large sample of normalization channel candidates.

\subsection{Control channel}\label{subsec:control}
Since the signal simulation does not reproduce the distributions in data perfectly, the change in efficiency due to the discrepancies needs to be estimated.
The agreement between data and simulation is studied by comparing background-subtracted data and simulation distributions in the control and normalisation channels with a twofold strategy. In a first step, weights are computed as the ratio of data and simulation in the control channel, and they are applied to the simulated samples. Then, the remaining discrepancies are measured on both the control and the normalisation channels. 

As the reconstructed muon and photon distributions in the normalisation channel are different to those in the signal channel, a 3-body decay is used as control channel to correct the simulation. The \Bsphig decay, where the \phiz decays into \Kp\Km, is therefore used to control the reconstructed variables such as isolation, vertex $\chi^2$ and muon and photon kinematic distributions. The branching fraction of this decay, $(3.4\pm0.4)\times10^{-5}$ \cite{PDG2022}, is more than three orders of magnitude larger than that of the signal, ensuring a large sample size to perform the comparisons.

\begin{figure}[tb]
    \centering
    \includegraphics[width=.55\textwidth]{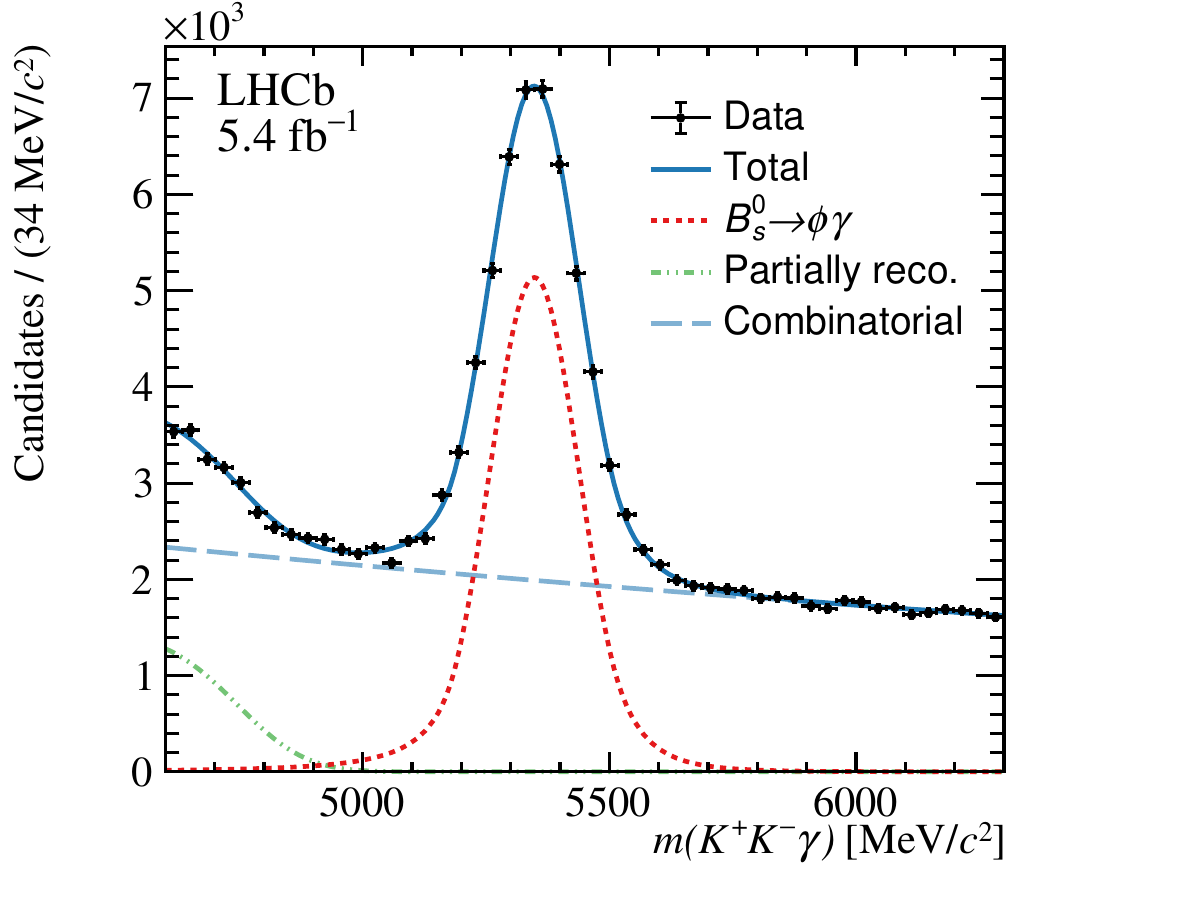}
    \caption{\small Mass distribution of \Bsphig candidates from data, used as control channel. The fit to the distribution is superimposed.}
    \label{fig:Bsphig}
\end{figure}

The \Bs candidate mass distribution is fitted with a double-sided Crystal Ball (DSCB) function~\cite{DSCB}, comprising a Gaussian core with asymmetric tails, with the tail parameters fixed to values obtained from simulation. Partially reconstructed backgrounds, such as the $\B\to\phiz\g\kaon$ decay, are described by an Argus function~\cite{argus} convolved with a Gaussian function. The combinatorial background contribution is  modelled by an exponential function. The resulting fit to the invariant-mass distribution of the $\Kp\Km\gamma$ candidates, is shown in Fig.~\ref{fig:Bsphig}. The fitted  \Bsphig yield is $36400 \pm 400$, where the uncertainty is statistical.

Background-subtracted distributions of variables are extracted using the \sPlot method~\cite{sPlot} and compared with simulation.
To avoid large weights from candidates coming from the tails of the signal distribution, where the ratio between background and signal is large, the mass window $[5.0,5.7]$\gevcc is used.

To find the optimal binning of the control variables, the \emph{GradBoost} weighting method~\cite{HepML} is used. To avoid a training bias, the control sample is randomly split into two parts and classifiers are trained on one part and tested on the other. The  weight is obtained as the average of the output of the two classifiers.

The weighting of the simulated samples is performed on three variables: the number of long tracks, the energy of the \Bs candidate, and its pseudorapidity. Weighting on the number of long tracks allows  the correction of the variables related to the event occupancy such as isolation and vertex $\chi^2$.
Correcting for the discrepancies in energy and pseudorapidity of the \Bs candidates, results in better matching of the kinematic variables of the decay products.

After the weighting procedure, the remaining discrepancies between data and simulation in the variables used to train the MLP  are calculated as the difference of the mean of the distributions and they range [1.4\textperthousand,~13$\%$].
These discrepancies are propagated through the second MLP classifier by shifting each input variables by its maximum discrepancy value. 
The change of efficiency of the shifted MLP with respect to the central one is considered as a systematic uncertainty. This is the dominant source of uncertainty, being 40\% of the total systematic uncertainty.


\section{Branching fraction determination}
\label{sec:BRdetermination}

The branching fraction of the signal is estimated by comparing its yield with a normalisation channel with well-known branching fractions:
$\Bs\to\jpsi\eta$, with $\jpsi\to\mup\mun$ and $\eta\to\g\g$. This also allows a partial cancellation in the ratio of efficiencies which reduces the total uncertainty on the final result.
The branching fraction is therefore expressed as \begin{equation}
    \label{eq:NormalizedBR}
    \BR(\Bsmmg) = \frac{\BR_\text{norm}}{N_\text{norm}} \times \dfrac{\epsilon_\text{norm}}{\epsilon_{\text{sig}}} \times N_{\text{sig}} \,,
\end{equation}
where $\epsilon_\text{norm}$ and $\epsilon_{\text{sig}}$ are the normalization and signal efficiencies, respectively. From the latest \CP-averaged values of $\BR(\Bsjpsie)$, $\BR(\jpsi\to\mup\mun)$, and $\BR(\eta\to\g\g)$, the combined branching fraction of the normalisation channel is \mbox{$\BR_\text{norm} = (9.3 \pm 1.6) \times 10^{-6}$}~\cite{PDG2022}.

\begin{figure}[tb]
    \centering
\includegraphics[width=.55\textwidth]{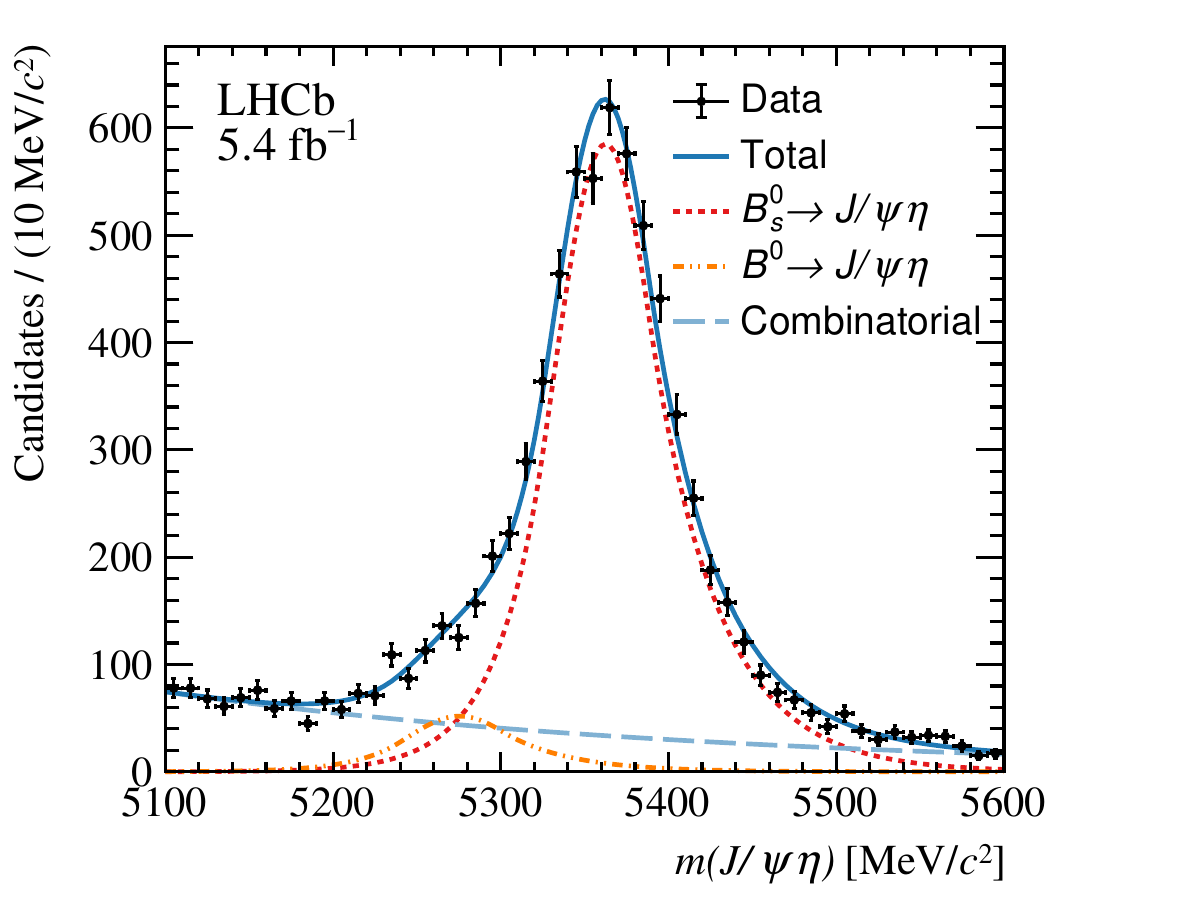}%
    \caption{Mass distribution of \Bsjpsie candidates from data, used as normalisation channel. The fit to the distribution is superimposed.}
    \label{fig:BsjpsieDirect}
\end{figure}

In order to estimate the \Bsjpsie yield, a fit to the data is performed accounting for contributions from $\Bzero\to\jpsi\etaz$ decays and combinatorial background. The \Bsjpsie candidates are modelled with a DSCB distribution with tail parameters fixed to values determined from simulation. The $\Bzero\to\jpsi\etaz$ contribution is modelled with the same shape used for the signal, with the mean shifted by the difference between the known \Bzero and \Bs masses, and the relative contribution left floating to account for the difference in efficiency. The combinatorial background is modelled with an exponential distribution with the slope parameter left floating.
 The invariant-mass distribution for the selected candidates of the normalisation channel, and the fit result, are shown in Fig.~\ref{fig:BsjpsieDirect}. The fitted \Bsjpsie yield is $N_{\mathrm{norm}} = 5680 \pm 110$, where the uncertainty is statistical.

\subsection{Efficiencies and their systematic uncertainties}
\label{subsec:effs}
The efficiencies to detect the signal and normalisation channels can be factorised as \begin{equation}
    \label{eq:eff}
    \epsilon =\epsilon_\text{Acceptance}\times
    \epsilon_\text{Preselection}\times
    \epsilon_{\mu \text{PID}}\times
    \epsilon_{\gamma \text{PID}}\times
    \epsilon_\text{Trigger}\times
    \epsilon_\text{MLP}.
\end{equation}
Each efficiency and its associated systematic uncertainty are evaluated relative to the preceding stage for the signal and normalisation channels. The acceptance efficiency is defined as the fraction of decays with daughters inside the detector acceptance.

The efficiency of the preselection described in Sec.~\ref{sec:selection} includes the efficiency of reconstructing all the final state particles. It is calculated from simulation for both signal and normalisation channels. A systematic uncertainty associated with the truth-matching algorithm described in Sec.~\ref{sec:Detector} is estimated by fitting the reconstructed \Bs mass distribution of simulated candidates failing this algorithm.

The muon identification efficiency, $\epsilon_{\mu \text{PID}}$, is measured using high-purity calibration samples of $\jpsi\to\mup\mun$ decays, selected using only kinematic requirements~\cite{LHCb-PUB-2016-021}.
The muon PID efficiency is evaluated as a function of the muon momentum and pseudorapidity, as well as the track multiplicity of the event using a dedicated procedure~\cite{LHCb-PUB-2016-021}. The resulting efficiency maps are applied to simulated samples to determine the integrated efficiency for the signal and normalisation channels. This method introduces two systematic uncertainties. The first one is associated to the background subtraction method used in the calibration samples and the second one 
from modeling the dependencies of the muon PID efficiency maps. These uncertainties are calculated for each data-taking year and they are in the range [6,~7]\textperthousand.

The photon identification efficiency, $\epsilon_{\gamma \text{PID}}$, is estimated from data with a similar method as the one applied to muons. The efficiency is extracted from a sample of $\eta\to\mup\mun\gamma$ decays, as a function of photon momentum, pseudorapidity and number of tracks in the event. The integrated efficiency is computed using the corrected signal kinematics for signal and normalisation channels. Similarly to the charged PID method, this method introduces two systematic uncertainties in the signal to normalisation efficiency ratio. The first  is associated to the fit and the background subtraction method in the $\eta\to\mup\mun\gamma$ decay. The second uncertainty is estimated from the modeling of the efficiency dependency.
These uncertainties are calculated for each data-taking year and are in the range [1,~5]\%.

The trigger efficiency, $\epsilon_\text{Trigger}$, is determined from data with the TISTOS method~\cite{LHCb-PUB-2014-039}. Trigger information is associated to the reconstructed candidates during the offline processing. A reconstructed signal candidate can be classified into three categories: candidates triggered on signal (TOS), triggered on parts of the underlying event that are independent of the tracks forming the signal candidate (TIS), or triggered on both elements of the signal candidate and the underlying event.
The trigger efficiency can be estimated by exploiting the overlap between the TIS and TOS categories (TIS$\&$TOS) and assuming that the signal decays are uncorrelated with the rest of the event. The trigger efficiency of a given decay channel, is computed from the number of triggered candidates, $N_\text{Trigger}$,  with respect to a total of $N_\text{Total}$ candidates as \begin{equation}
    \label{eq:TISTOS}
    \epsilon_\text{Trigger} = \frac{N_\text{Trigger}}{N_\text{Total}} = 
    \frac{N_\text{Trigger}}{N_\text{TIS}}\times \epsilon_\text{TIS} =
    \frac{N_\text{Trigger}}{N_\text{TIS}} \frac{N_\text{TIS\&TOS}}{N_\text{TOS}} \, ,
\end{equation}
where $N_i$ is the number of background-subtracted candidates triggered within the category~$i$.
The \Bsphig candidates are used to calculate photon trigger efficiencies, while muon and TIS trigger efficiencies are calculated from $B^{+}\to\jpsi K^{+}$ candidates.
The signal trigger efficiencies are calculated by convolving the simulated signal kinematics on the following variables: $p_z(B)$ and $\pt(B)$ for TIS trigger,  $\pt(\mu)$ and IP($\mu$) for the muon triggers, and $\pt(\gamma)$ for photon triggers. The systematic uncertainties of the trigger efficiencies are estimated by determining them with a different method that makes use of simulation. The efficiency ratios differ between the two methods by [2,~3]\%, which is assigned as systematic uncertainty.

Finally, the efficiency of the first and second MLP requirements, $\epsilon_\text{MLP}$, as well as their associated systematic uncertainties, are calculated for the signal and normalisation channels using the kinematically weighted simulation presented in Sec.~\ref{subsec:control}.

The ratio of efficiencies for each $q^2$ region and the single-event sensitivity, $\alpha$, defined as
\begin{equation}
    \alpha \equiv \frac{\BR_\textrm{norm}}{N_\textrm{norm}}\times \dfrac{\epsilon_\text{norm}}{\epsilon_{\text{sig}}},
\label{eq:sensitivity}
\end{equation}
are shown in Table~\ref{tab:normfactors_singleeventsens}, for each $q^2$ bin.

\begin{table}[htb]
    \small
    \centering \caption{Efficiency ratios and single-event sensitivities, for each $q^2$ bin. The uncertainties on the efficiency ratios include the systematic uncertainties presented in Sec.~\ref{subsec:effs}. The uncertainties on $\alpha$ are the propagated uncertainties from  $\BR_\textrm{norm}$, $N_\textrm{norm}$ and $\epsilon_\text{norm}/\epsilon_{\text{sig}}$.}
    \begin{tabular}{cccc} \hline \hline
         $q^2$ bin & I & II & III\\ \hline
         $\epsilon_\text{norm}/\epsilon_{\text{sig}}$&  0.85 $\pm$ 0.07 & 0.95 $\pm$ 0.08 & 2.20 $\pm$ 0.07\\
         $\alpha [10^{-9}]$&  1.40 $\pm$ 0.27& 1.56 $\pm$ 0.30 & 3.60 $\pm$ 0.63 \\
         \hline\hline
    \end{tabular}
    \label{tab:normfactors_singleeventsens}
\end{table}

\subsection{Signal and background model}
\label{sec:model}

The signal yield is determined from fits to the invariant-mass distribution of the selected candidates.
The signal model is extracted from simulation, and described with a DSCB function. 

The background contamination can be divided into four categories: combinatorial background, partially reconstructed decays, peaking and misidentified backgrounds. Combinatorial background candidates are distributed across the entire mass search region, peaking background candidates have a reconstructed mass close to the \Bs mass value, while partially reconstructed, and misidentified contributions populate the region below the \Bs mass. 

The combinatorial background is modelled using an exponential function with the slope and yield parameters left free to float in the fit. Processes where one particle of the final state is not reconstructed, referred to as partially reconstructed background, contribute with a broad peak below the \Bs mass. Three decays are studied in simulation: $\Bs\to\mup\mun\phi$ with $\phi\to\pip\pim\piz$, $\Bz\to\mup\mun\eta'$ with $\eta'\to\rhoz\gamma$, and 
$\Bz\to\Dm\mup\nu_\mu$  with $\Dm\to\piz\mun\neumb$. The invariant-mass distributions of these three decays are fitted with Argus functions. To account for the energy resolution, the Argus distribution is convolved with a Gaussian function centered at zero. Since the value of the Argus cutoff parameter has to be consistent among the three $q^2$ bins, it is fixed  to the average of the fitted values from simulation, on each decay and each $q^2$ region. The impact of fixing this parameter  is evaluated using simulated pseudoexperiments, by comparing the baseline result with alternative fits in which the different cutoff parameters are fixed to the individual values or a single cutoff parameter is allowed to float. The difference in the fitted signal yield is considered as systematic uncertainty. The yield and other model parameters are left floating in the fit.

The main background processes contributing to the signal mass region, referred to as peaking backgrounds, are the \Bmmpi decay with $\piz\to\g\g$ and the $B_{(s)}^0\to\mup\mun\eta$ decay with $\eta\to\g\g$. These decays contribute to the background if one photon is not reconstructed or if both photons are reconstructed in the same calorimeter cluster. The mass distributions of these decays are extracted from simulation, and the shape parameters are fixed in the final mass fit. The \Bmmpi decay is modelled by a DSCB function while the $B_{(s)}^0\to\mup\mun\eta$ is found to be well described by an Argus shape convolved with a Gaussian function. Their yields are estimated from the expected branching fractions in Ref.~\cite{PDG2022}, and the efficiencies determined from simulation. In the final mass fit, the yield parameters are Gaussian-constrained to the estimated values.


Apart from a slightly smaller mass and the CKM suppression, \Bmmg decays differ from their \Bs counterpart by their light meson resonances.
As discussed in Ref.~\cite{MelikhovNikitin_2004}, $\Bz\to\g$ transition form factors involve the $\rho$ and the $\omega$ resonances, while  $\Bs\to\g$ transition form factors involve the \phiz resonance, all of them in the low-$q^2$ region (bin~I) \cite{PDG2022}. Assuming that the branching fraction in bin~I is dominated by these resonances, one can approximate the number of candidates under the narrow-width approximation \cite{BsToMuMuGamma_Meril,PDG2022} \begin{align}
    \left.\frac{N_{\Bz\to\mup\mun\g}}{N_{\Bs\to\mup\mun\g}} \right|_\text{I} &\approx
    \frac{f_d}{f_s}\,
\frac{\BR(\Bz\to\rho\g)\BR(\rho\to\mup\mun)+\BR(\Bz\to\omega\g)\BR(\omega\to\mup\mun)}{\BR(\Bs\to\phi\g)\BR(\phi\to\mup\mun)}\nonumber\approx0.03,
\end{align}
where $f_d$ and $f_s$ are the fragmentation fractions~\cite{PDG2022}.
In the two other $q^2$ regions, II and III, where there are not resonant contributions, the relative contribution of \Bz with respect to \Bs can be estimated as \begin{align}
    \left.\frac{N_{\Bz\to\mup\mun\g}}{N_{\Bs\to\mup\mun\g}} \right|_\text{II}  \approx
    \left.\frac{N_{\Bz\to\mup\mun\g}}{N_{\Bs\to\mup\mun\g}} \right|_\text{III}  \approx
    \frac{f_d}{f_s} \left|\frac{\Vtd}{\Vts}\right|^2 \approx 0.17 \,,\nonumber
\end{align}
where $\Vtd$ and $\Vts$ are the CKM matrix elements. The \Bmmg contribution is therefore considered negligible within the signal statistical uncertainty in all the $q^2$ regions.

Physical backgrounds can also emerge from doubly misidentified hadronic decays. The main ones are \BsPhiGam decays where $\phiz\to\Kp\Km$ and \BKstGam decays where $\Kstarz \to\Kp\pim$. The mis-identification probabilities are determined as a function of the track momentum and transverse momentum, using $D^{*+}\to \Dz\pip$ decays with $\Dz\to\Km\pip$ with the method described in Ref.~\cite{LHCb-PAPER-2021-008}, and their contributions are found to be negligible.
The effect of neglecting these two contributions and the \Bmmg contamination is studied. The CL$_\mathrm{S}$~\cite{CLs} method is applied to determine branching fraction limits from simulated pseudoexperiments generated from the  mass sidebands under the background-only hypothesis, with and without including the three backgrounds in the fit. The two results differ by  less than one percent, confirming that their contributions are negligible.

Other possible backgrounds were studied and found to be negligible, such as $\Lambda_b$ decays, charmonium resonances decaying to muons and $\Bz\to\pip\pim\piz$ decays with double pion misidentification and a merged \piz.

The possibility of contamination from the $B^{*0}\to B^{0}\gamma$ decay with $B^{0}\to\pip\pim$ where both pions are misidentified as muons, is studied. The muon mass hypothesis is replaced by the pion mass, and the invariant-mass spectrum is analysed. No candidates with reconstructed $m(\pip\pim)$ in the \Bz mass region are seen, therefore, the contribution from the $B^{*0}\to B^{0}\gamma$ decay is found to be negligible.

\begin{figure}[ht!]
    \centering
    \includegraphics[width=0.53\textwidth]{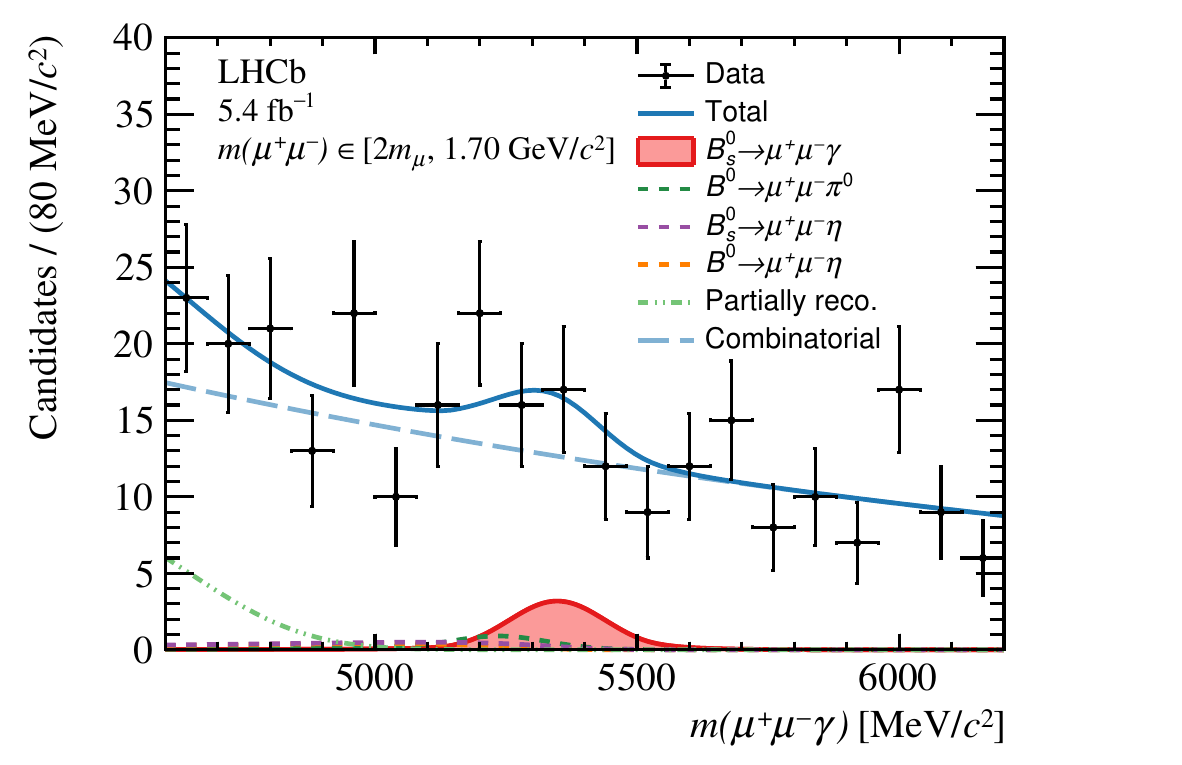}\includegraphics[width=0.53\textwidth]{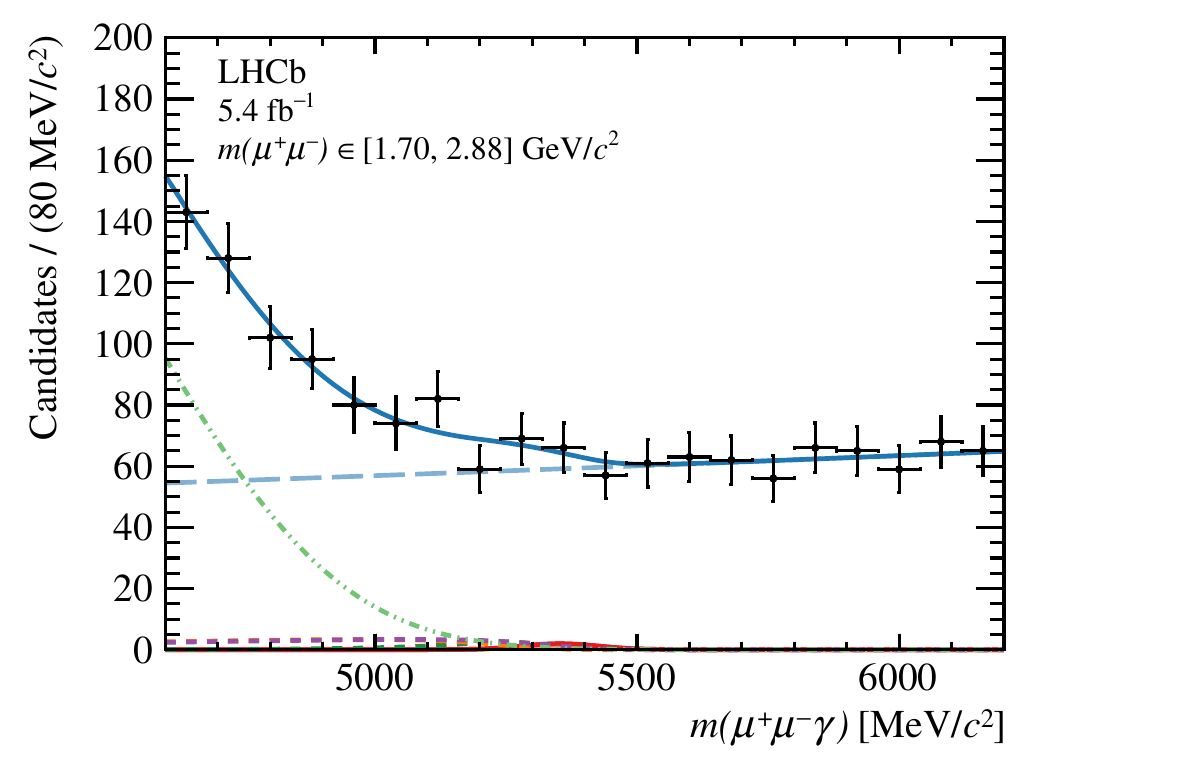}
    \includegraphics[width=0.53\textwidth]{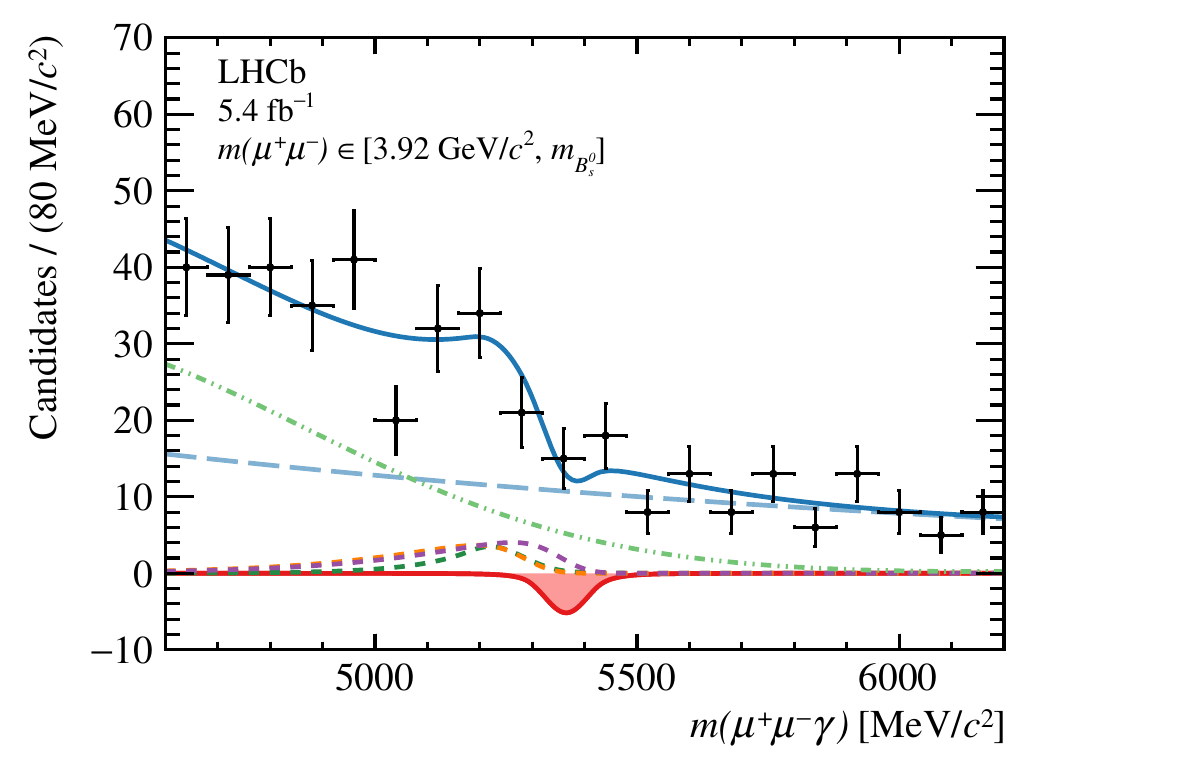}\includegraphics[width=0.53\textwidth]{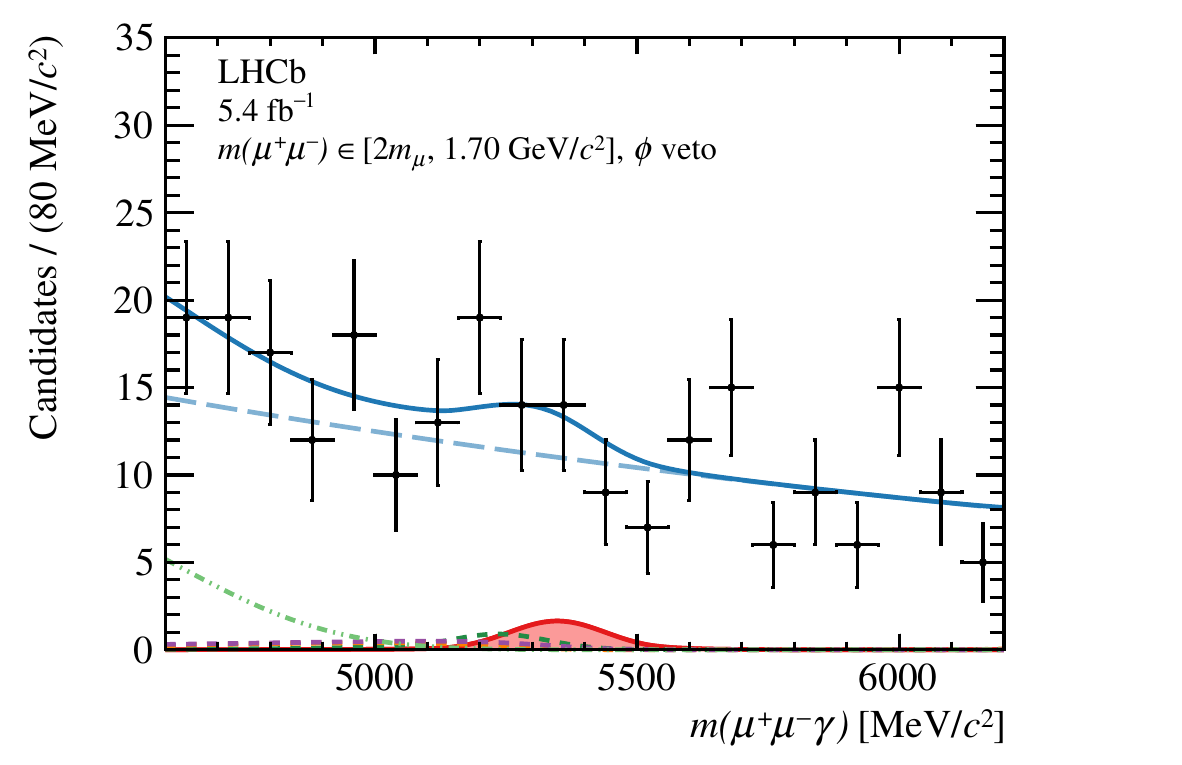}
    \caption{Mass distribution of signal candidates in regions of $q^2$, (top-left) bin~I, (top-right) bin~II,  (bottom-left) bin~III, and (bottom-right) \phiz-veto bin. The result of the fit is overlaid and the different components detailed in the legend. }
    \label{fig:FinalMass}
\end{figure}

The fit is performed in the invariant-mass range [4.60,~6.20]\gevcc. The invariant-mass distribution is shown in Fig.~\ref{fig:FinalMass} for all $q^2$ bins. The low-mass region is populated by partially reconstructed background, while the higher mass region is dominated by combinatorial background. The sources of systematic uncertainties introduced in Secs.~\ref{sec:selection}--\ref{subsec:effs} are included in the fit as nuisance parameters, where the neutral PID and the differences between data and simulation are the dominant sources.

\section{Results}
\label{sec:results}

The branching fraction of the signal decay is determined using an unbinned extended maximum-likelihood fit to the invariant-mass distribution of the contributions. The resulting branching fractions are\begin{eqnarray}\label{eq:BRvalues}
   \BR(\Bsmmg)_\textrm{I}  &=& (1.34\pm1.60\pm0.28 )\times10^{-8},\nonumber\\
   \BR(\Bsmmg)_\textrm{II}  &=& (0.76\pm3.55\pm0.30 )\times10^{-8},\nonumber\\
   \BR(\Bsmmg)_\textrm{III}  &=& (-2.55\pm2.25\pm0.41 )\times10^{-8},\nonumber\\
   \BR(\Bsmmg)_\textrm{I, with \phiz veto}  &=& (0.72\pm1.56\pm0.29)\times10^{-8}.\nonumber
\end{eqnarray}
The statistical uncertainties are evaluated by repeating the fits with all nuisance parameters fixed to the values obtained in the standard fit, where all nuisance parameters are free to float within their constraints. The systematic uncertainties are  computed by subtracting the statistical uncertainties in quadrature from those obtained with the nuisance parameters floating.

The signal branching fractions are consistent with zero in all $q^2$ regions and consistent with the background-only hypothesis at the $<1\sigma$ level. Limits are set on $\BR(\Bsmmg)$  using the CL$_{\mathrm{s}}$ method~\cite{CLs}  with a one-sided test statistic~\cite{CLside} as implemented in Refs.~\cite{GammaCombo,LHCb-PAPER-2021-033,LHCb-PAPER-2016-032}. The one-sided test statistic for a given branching fraction value is defined as twice the negative logarithm of the profile likelihood ratio if it is larger than the measured branching fraction and zero otherwise. Its distribution is determined from pseudoexperiments, where nuisance parameters are set to their best fit values when generating the samples, while central values of the Gaussian-constraints are independently fluctuated within their uncertainty for each pseudoexperiment as described in Ref.~\cite{CLs_Nuisance}.

\begin{figure}[ht!]
    \centering
    \includegraphics[width=0.5\textwidth]{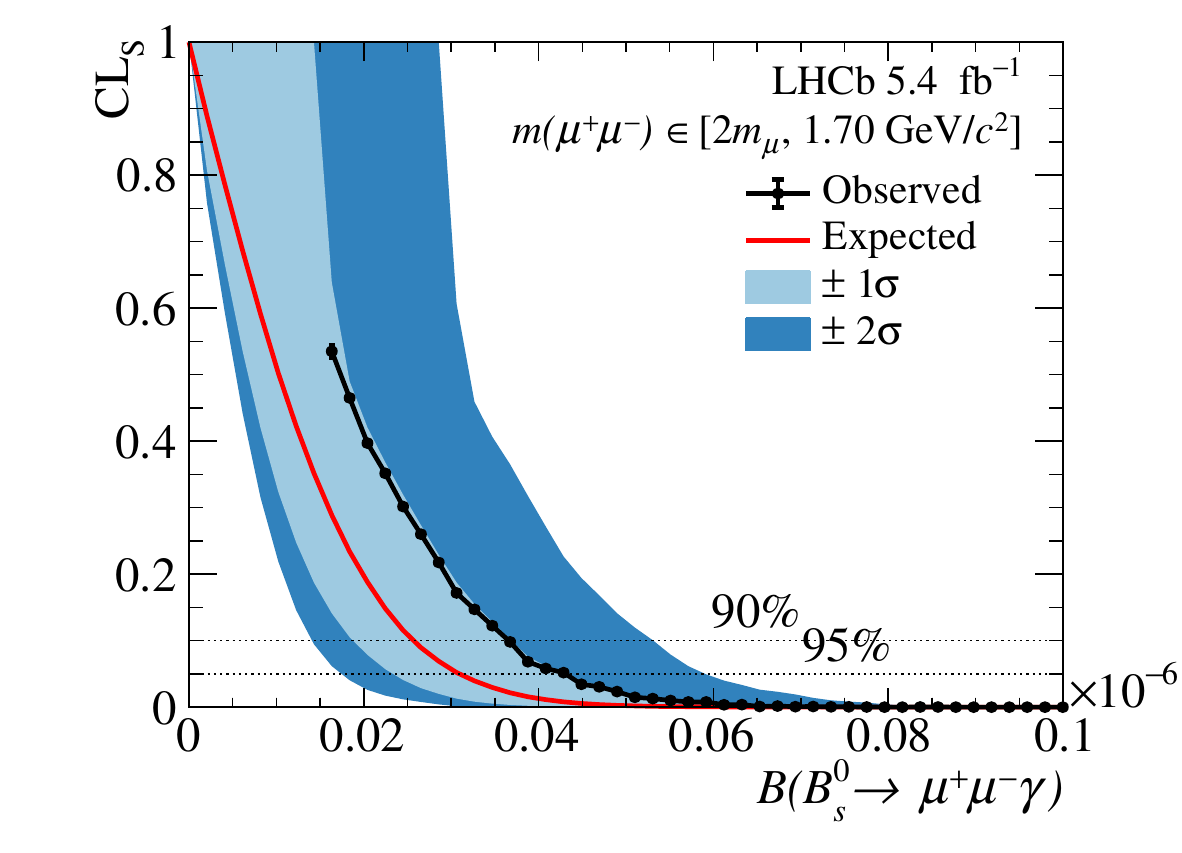}\includegraphics[width=0.5\textwidth]{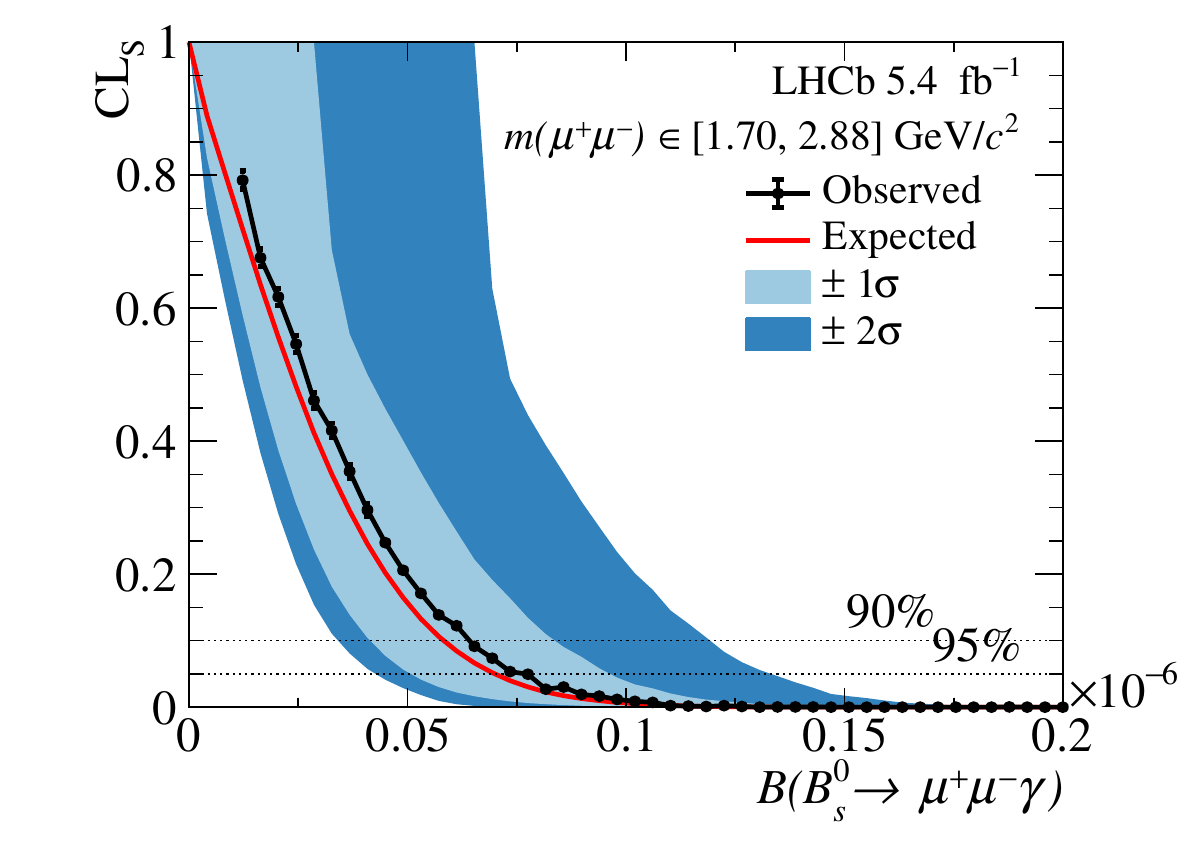}\\
\includegraphics[width=0.5\textwidth]{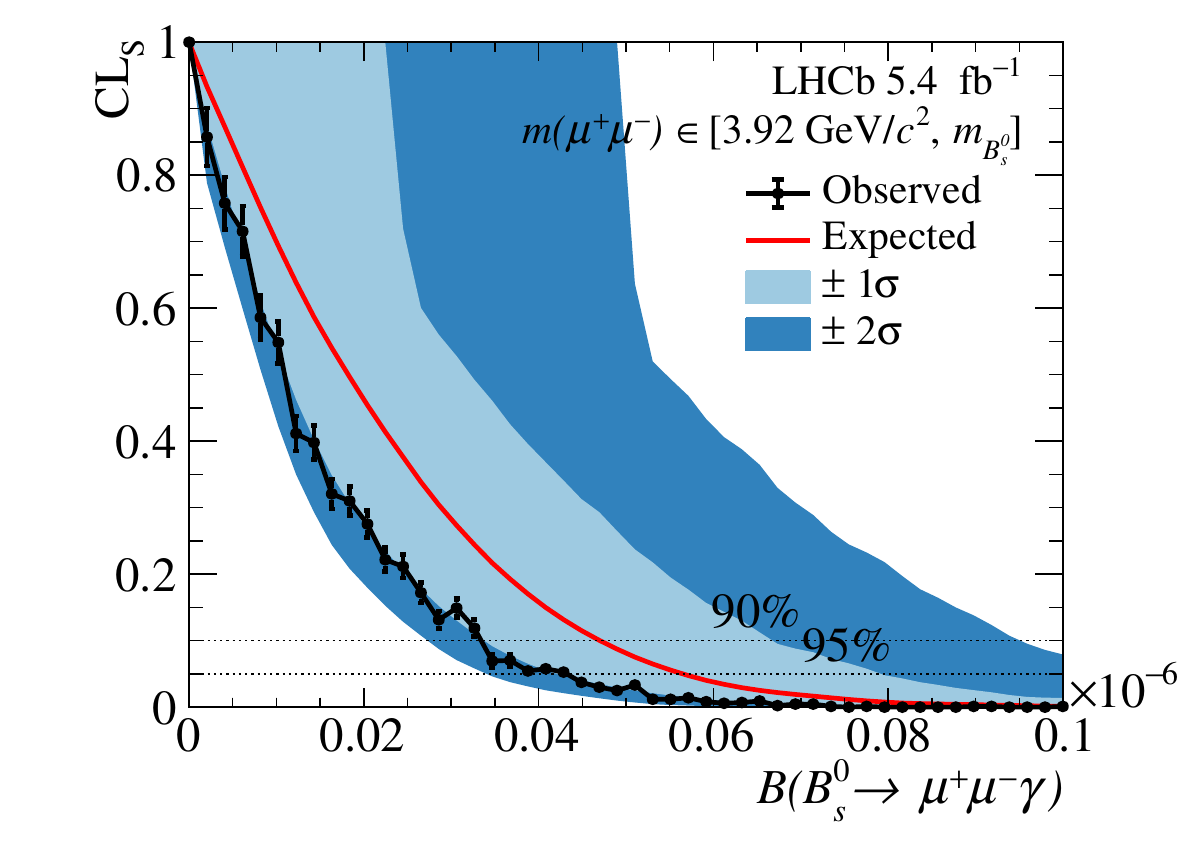}\includegraphics[width=0.5\textwidth]{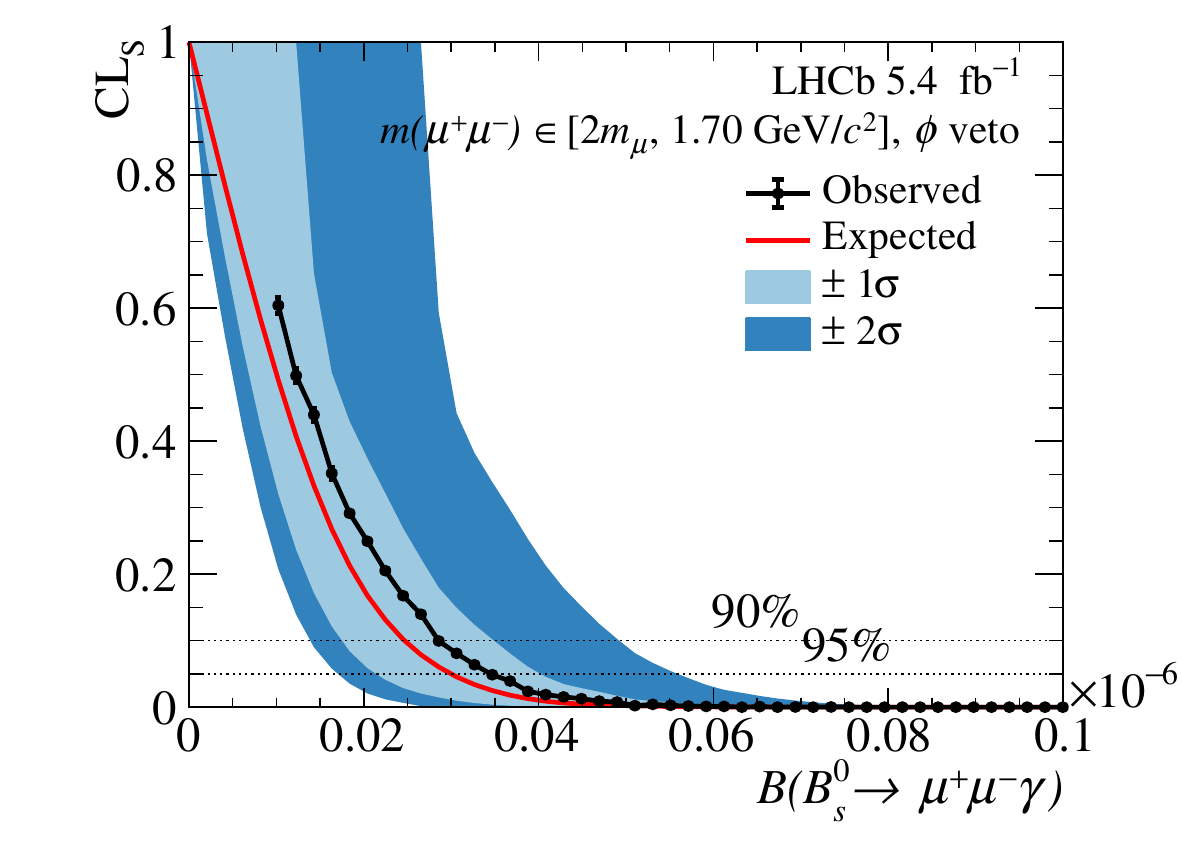}\\
\includegraphics[width=0.5\textwidth]{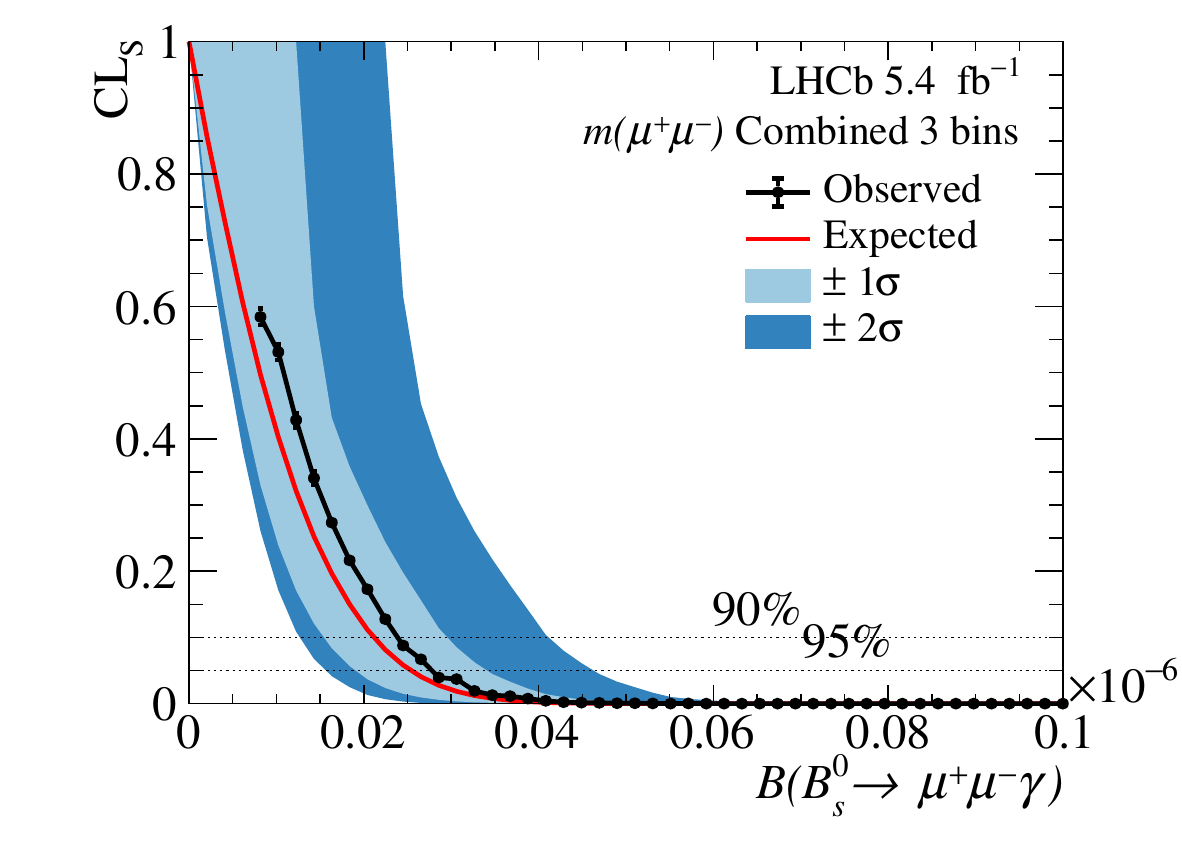}
    \caption{Results from the CL$_{\mathrm{s}}$ scan used to obtain the limit on $\BR(\Bsmmg)$ in regions of $q^2$: (top-left) bin~I, (top-right) bin~II, (middle-left) bin~III, (middle-right) bin~I with \phiz veto, and (bottom) the whole $q^2$ range. The background-only expectation is shown by the red line and the 1$\sigma$ and 2$\sigma$ bands are shown as light blue and blue bands, respectively. The observed limit is shown as the solid black line. The two dashed lines intersecting with the observation indicate the limits at 90\% and 95\% CL for the upper and lower line, respectively. }
    \label{fig:gammacombo_UB}
\end{figure}

The fit is also performed simultaneously in the three $q^2$ regions to obtain a combined branching fraction limit.
The CL$_{\mathrm{s}}$ curves are shown in Fig.~\ref{fig:gammacombo_UB} from which the limit on the branching fraction for each $q^2$ region is found to be \begin{eqnarray}\label{eq:BRlimits}
    \BR(\Bsmmg)_\textrm{I} < 3.6\,(4.2)\times10^{-8},&\nonumber\\
    \BR(\Bsmmg)_\textrm{II} < 6.5\,(7.7)\times10^{-8},&\nonumber\\
    \BR(\Bsmmg)_\textrm{III} < 3.4\,(4.2)\times10^{-8},&\nonumber\\
    \BR(\Bsmmg)_\textrm{I, with \phiz veto} < 2.9\,(3.4)\times10^{-8},&\nonumber\\ [8pt]
    \BR(\Bsmmg)_\textrm{comb.} < 2.5\,(2.8)\times10^{-8},&\nonumber
\end{eqnarray}
at 90\%~(95\%) CL. The observed upper limits are shown in Fig.~\ref{fig:gammacombo_UB}, together with the background-only expectations.

\begin{figure}[ht!]
    \centering
    \includegraphics[width=0.85\textwidth]{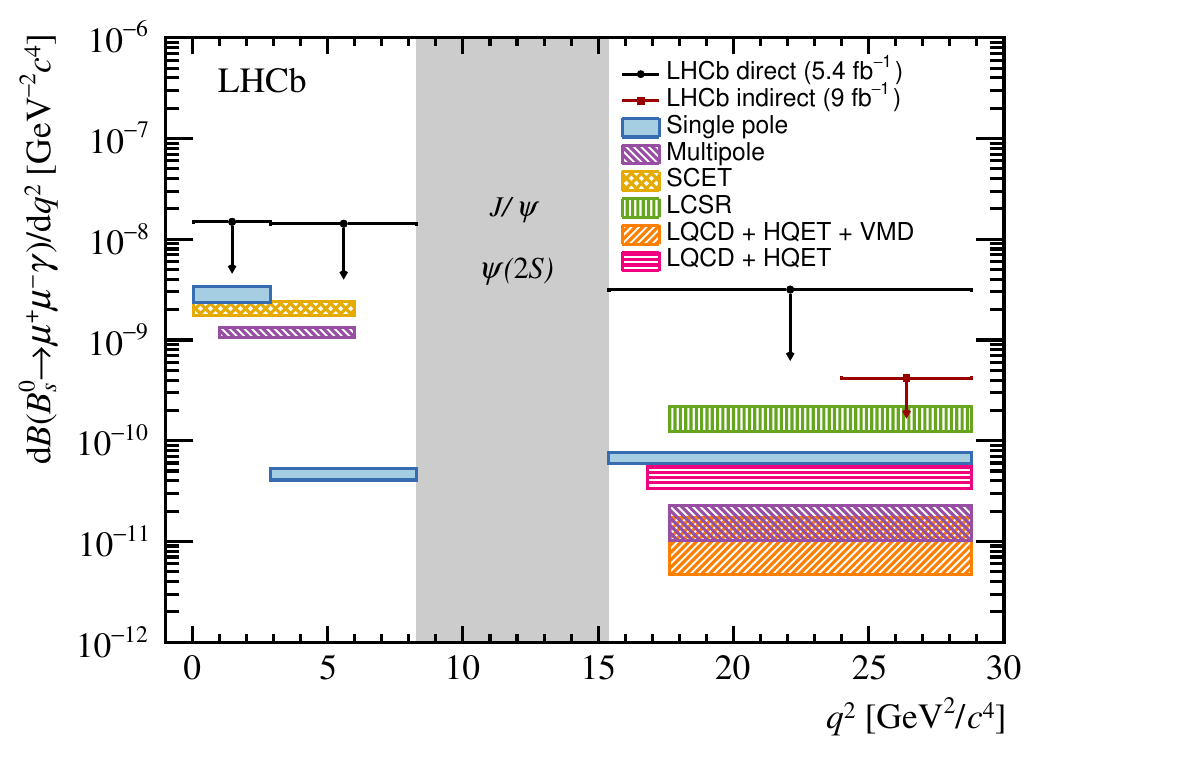}
    \caption{Differential branching fraction of the \Bsmmg decay in intervals of $q^2$. The black arrows represent the limits set in this paper at 95\% CL. The limit set in the high-$q^2$ region with the indirect method from the \Bsmm measurement at 95\% CL is represented in red~\cite{LHCb-PAPER-2021-008}.
   SM predictions using different form factor calculations are represented in 
(blue) Ref.~\cite{BsToMuMuGamma_Meril}, 
(violet) Ref.~\cite{Nikitin_2018},
(yellow) Ref.~\cite{Beneke_2020},
(green) Ref.~\cite{Janowski_2021}, 
(orange) Ref.~\cite{Bsmumugamma_Camille,Bsmumugamma_Camille2}, and (pink) Ref.~\cite{frezzotti_2024}.
}
    \label{fig:ResultPlot}
\end{figure}

The limits are shown in Fig.~\ref{fig:ResultPlot}, together with the limit set by the indirect method~\cite{LHCb-PAPER-2021-008} for the $m(\mup\mun)$ range [4.9,~$m_{\Bs}$]\gevcc. They are consistent with all SM predictions. The different theoretical approaches to the local form factor calculations predict different branching fractions, which are also shown in Fig.~\ref{fig:ResultPlot} as coloured bands. The theoretical predictions shown are calculated by a 
single-pole~\cite{BsToMuMuGamma_Meril} 
and multipole~\cite{Nikitin_2018} parametrisation,
using SCET~\cite{Beneke_2020}, 
from LCSR~\cite{Janowski_2021},  
and from lattice QCD using HQET extrapolation~\cite{frezzotti_2024}
and assuming VMD~\cite{Bsmumugamma_Camille,Bsmumugamma_Camille2}. 

\section{Conclusions}
\label{sec:conclusions}

In summary,  data collected with the LHCb detector in the years 2016\,--\,2018, corresponding to an integrated luminosity of 5.4\,fb$^{-1}$ of proton-proton ($pp$) collisions at a center-of-mass energy $\sqrt{s}=13$\tev are analysed to search for the \Bsmmg decay. The search is performed in four regions of the dimuon invariant mass.

The measured signal is not statistically significant, and consistent with the background-only hypothesis at less than $1\sigma$ level in all dimuon mass regions. Upper limits on the branching fractions are set at 
 \begin{align}
    \BR(\Bsmmg) < 4.2\times10^{-8},~&m(\mup\mun)\in[2m_\Pmu,~1.70]\gevcc,\nonumber\\
    \BR(\Bsmmg) < 7.7\times10^{-8},~&m(\mup\mun)\in[1.70,~2.88]\gevcc,\nonumber\\
    \BR(\Bsmmg) < 4.2\times10^{-8},~&m(\mup\mun)\in[3.92,~m_\Bs]\gevcc,\nonumber
    \end{align} and \begin{align}
    \BR(\Bsmmg) < 2.8\times10^{-8}\nonumber
\end{align}
for the combined dimuon mass regions, at 95\% CL.  Additionally, an upper limits are set at $\BR(\Bsmmg) < 3.4\times10^{-8}$ in the $[2m_\Pmu,~1.70]\gevcc$ dimuon mass region excluding the contribution from the intermediate \phiz meson. These are the first limits set on the \Bsmmg decay with full final state reconstruction and the first limit at dimuon masses below 4.9\gevcc.

\section*{Acknowledgements}
%
%
\noindent We express our gratitude to our colleagues in the CERN
accelerator departments for the excellent performance of the LHC. We
thank the technical and administrative staff at the LHCb
institutes.
We acknowledge support from CERN and from the national agencies:
CAPES, CNPq, FAPERJ and FINEP (Brazil); 
MOST and NSFC (China); 
CNRS/IN2P3 (France); 
BMBF, DFG and MPG (Germany); 
INFN (Italy); 
NWO (Netherlands); 
MNiSW and NCN (Poland); 
MCID/IFA (Romania); 
MICINN (Spain); 
SNSF and SER (Switzerland); 
NASU (Ukraine); 
STFC (United Kingdom); 
DOE NP and NSF (USA).
We acknowledge the computing resources that are provided by CERN, IN2P3
(France), KIT and DESY (Germany), INFN (Italy), SURF (Netherlands),
PIC (Spain), GridPP (United Kingdom), 
CSCS (Switzerland), IFIN-HH (Romania), CBPF (Brazil),
and Polish WLCG (Poland).
We are indebted to the communities behind the multiple open-source
software packages on which we depend.
Individual groups or members have received support from
ARC and ARDC (Australia);
Key Research Program of Frontier Sciences of CAS, CAS PIFI, CAS CCEPP, 
Fundamental Research Funds for the Central Universities, 
and Sci. \& Tech. Program of Guangzhou (China);
Minciencias (Colombia);
EPLANET, Marie Sk\l{}odowska-Curie Actions, ERC and NextGenerationEU (European Union);
A*MIDEX, ANR, IPhU and Labex P2IO, and R\'{e}gion Auvergne-Rh\^{o}ne-Alpes (France);
AvH Foundation (Germany);
ICSC (Italy); 
GVA, XuntaGal, GENCAT, Inditex, InTalent and Prog.~Atracci\'on Talento, CM (Spain);
SRC (Sweden);
the Leverhulme Trust, the Royal Society
 and UKRI (United Kingdom).

\addcontentsline{toc}{section}{References}
\bibliographystyle{LHCb}
\bibliography{main,standard,LHCb-PAPER,LHCb-CONF,LHCb-DP,LHCb-TDR}

\newpage
\centerline
{\large\bf LHCb collaboration}
\begin
{flushleft}
\small
R.~Aaij$^{35}$\lhcborcid{0000-0003-0533-1952},
A.S.W.~Abdelmotteleb$^{54}$\lhcborcid{0000-0001-7905-0542},
C.~Abellan~Beteta$^{48}$,
F.~Abudin{\'e}n$^{54}$\lhcborcid{0000-0002-6737-3528},
T.~Ackernley$^{58}$\lhcborcid{0000-0002-5951-3498},
A. A. ~Adefisoye$^{66}$\lhcborcid{0000-0003-2448-1550},
B.~Adeva$^{44}$\lhcborcid{0000-0001-9756-3712},
M.~Adinolfi$^{52}$\lhcborcid{0000-0002-1326-1264},
P.~Adlarson$^{79}$\lhcborcid{0000-0001-6280-3851},
C.~Agapopoulou$^{46}$\lhcborcid{0000-0002-2368-0147}
% ...

{\footnotesize \it

% ...
$^{35}$Nikhef National Institute for Subatomic Physics, Amsterdam, Netherlands\\
% ...
$^{44}$Instituto Galego de F{\'\i}sica de Altas Enerx{\'\i}as (IGFAE), Universidade de Santiago de Compostela, Santiago de Compostela, Spain\\
% ...
$^{46}$European Organization for Nuclear Research (CERN), Geneva, Switzerland\\
% ...
$^{48}$Physik-Institut, Universit{\"a}t Z{\"u}rich, Z{\"u}rich, Switzerland\\
% ...
$^{52}$H.H. Wills Physics Laboratory, University of Bristol, Bristol, United Kingdom\\
% ...
$^{54}$Department of Physics, University of Warwick, Coventry, United Kingdom\\
% ...
$^{57}$School of Physics and Astronomy, University of Glasgow, Glasgow, United Kingdom\\
$^{58}$Oliver Lodge Laboratory, University of Liverpool, Liverpool, United Kingdom\\
% ...
$^{66}$Syracuse University, Syracuse, NY, United States\\
% ...
$^{79}$Department of Physics and Astronomy, Uppsala University, Uppsala, Sweden, associated to $^{57}$\\
% ...
}
\end{flushleft}

\end{document}